\documentclass[
    ,final            
  ]
  {aipproc}

\layoutstyle{8x11single}

\usepackage{placeins}
\usepackage{color}

\begin{document}
\newcommand{\gevc}{(GeV/c)$^2$~}
\newcommand{\gevcp}{(GeV/c)$^2$}
\newcommand{\abs}[1]{\vert#1\vert2}
\newcommand{\f}[1]{F_{#1}}
\newcommand{\fb}[1]{\overline{F_{#1}}}
\newcommand{\conj}[1]{#1^*}
\newcommand{\mlp}{M_{1+}}
\newcommand{\elp}{E_{1+}}
\newcommand{\llp}{L_{1+}}
\newcommand{\mlm}{M_{1-}}
\newcommand{\llm}{L_{1-}}
\newcommand{\eOp}{E_{0+}}
\newcommand{\lOp}{L_{0+}}

  \newcommand{\onehalf}{\mbox{$\frac{1}{2}$}}
  \newcommand{\threehalfs}{\mbox{$\frac{3}{2}$}}
  \newcommand{\threehalves}{\mbox{$\frac{3}{2}$}}
  \newcommand{\smfrac}[2]{\mbox{$\frac{#1}{#2}$}}
%
  \newcommand{\Mo}{\mbox{$M1$}}
  \newcommand{\Ct}{\mbox{$C2$}}
  \newcommand{\Et}{\mbox{$E2$}}
  \newcommand{\Mop}{\mbox{$M_{1+}$}}
  \newcommand{\Mom}{\mbox{$M_{1-}$}}
  \newcommand{\Mopc}{\mbox{$M_{1+}^*$}}
  \newcommand{\Momc}{\mbox{$M_{1+}^*$}}
  \newcommand{\Eop}{\mbox{$E_{1+}$}}
  \newcommand{\Eom}{\mbox{$E_{1-}$}}
  \newcommand{\Ezp}{\mbox{$E_{0+}$}}
  \newcommand{\Eopc}{\mbox{$E_{1+}^*$}}
  \newcommand{\Eomc}{\mbox{$E_{1-}^*$}}
  \newcommand{\Ezpc}{\mbox{$E_{0+}^*$}}
  \newcommand{\Sop}{\mbox{$S_{1+}$}}
  \newcommand{\Som}{\mbox{$S_{1-}$}}
  \newcommand{\Szp}{\mbox{$S_{0+}$}}
  \newcommand{\Sopc}{\mbox{$S_{1+}^*$}}
  \newcommand{\Somc}{\mbox{$S_{1-}^*$}}
  \newcommand{\Szpc}{\mbox{$S_{0+}^*$}}

\newcommand{\GNdelta}{$\gamma N \rightarrow \Delta~$}
\newcommand{\Deltap}{\mbox{$\Delta^+$}}
\newcommand{\pn}{\mbox{$P_n$}}
\newcommand{\thtpq}{\mbox{$\theta_{pq}^*$}}
\newcommand{\phipq}{\mbox{$\phi_{pq}$}}
\newcommand{\Ndelta}{$N \rightarrow \Delta~$}

\newcommand{\RLT}{\mbox{$R_{LT}$}}
\newcommand{\RLTn}{\mbox{$R_{LT}^n$}}
\newcommand{\RT}{\mbox{$R_T$}}
\newcommand{\RTT}{\mbox{$R_{TT}$}}

\newcommand{\sLT}{\mbox{$\sigma_{LT}$}}
\newcommand{\sLTprime}{\mbox{$\sigma_{LT}^{\prime}$}}
\newcommand{\sLTn}{\mbox{$\sigma _{LT}^n$}}
\newcommand{\sT}{\mbox{$\sigma_T$}}
\newcommand{\sTT}{\mbox{$\sigma_{TT}$}}
\newcommand{\sEtwo}{\mbox{$\sigma_{E2}$}}

\newcommand{\ndelta}{\ifmmode {N\!\!\rightarrow\!\!\Delta}
\else{$N\!\!\rightarrow\!\!\Delta$}\fi}
\newcommand{\alt}{\ifmmode {{\rm A}_{\rm LT}} \else {${\rm A}_{\rm LT}$}\fi}

\title{Overview: The Shape of Hadrons\footnote{This is the introductory article for the volume on the Shape of Hadrons, 
 Athens, Greece, 27-29 April 2006
AIP Conference Proceedings, Volume 904(2007), C.N. Papanicolas, and A.M. Bernstein, editors. }}

\newcommand{\mitlns}{Department of Physics and  Laboratory
for Nuclear Science\\ Massachusetts Institute of Technology\\
Cambridge, Massachusetts 02139, USA}

\newcommand{\athens}{Institute of Accelerating Systems and Applications, Athens, Greece \\
        and Department of Physics, University of Athens, Greece}

\author{A. M. Bernstein}{
  address={\mitlns}
}

\author{C.N. Papanicolas}{
  address={\athens}
}


\begin{abstract} 
In this article we address the physical basis of the deviation
of hadron shapes from spherical symmetry (non-spherical
amplitudes)  with  focus on the nucleon and $\Delta$. An overview
of both the experimental methods and results and the current
theoretical understanding of the issue is presented. At the
present time the most quantitative method is the $\gamma^{*} p
\rightarrow \Delta$ reaction for which significant non-spherical
electric (E2) and Coulomb quadrupole (C2) amplitudes have been
observed with good precision as a function of $Q^{2}$ from the
photon point through 6 GeV$^{2}$. Quark model calculations for
these quadrupole amplitudes are at least an order of magnitude too
small and even have the wrong sign. Lattice QCD, chiral effective
field theory, and dynamic model calculations which include the
effects of the pion-cloud are in approximate agreement with
experiment. This is expected due to the spontaneous breaking of
chiral symmetry in QCD and  the resulting, long range (low
$Q^{2}$) effects of the pion-cloud. Other observables such as
nucleon form factors and virtual Compton scattering experiments
indicate that  the pion-cloud is playing a significant role in
nucleon structure. Semi-inclusive deep inelastic scattering
experiments with transverse polarized beam and target also show
the effect of non-zero quark angular momentum. 

\end{abstract}

\maketitle 
\section{Preface}
This volume contains papers addressing issues of relevance to the
broad topic of the "Shape of Hadrons". These issues were examined
and debated in two meetings, which from the outset were planned to
be two phases of the same workshop. The first was held at MIT in
Aug. 7-9, 2004 and the second at the University of Athens  in
April 27-29, 2006. Many  issues were raised at the first workshop
  and  addressed at the second. The contributions contained in
this volume are from the second meeting. Events did not
completely conform to the original plan. Progress was more rapid
due to the completion of Chiral Effective Field Theory
calculations and the emergence of new, accurate  data. Due to
this the  old questions  transform themselves into new ones. At
the MIT workshop an important topic was whether the effects of
the pion cloud were being observed in the $\gamma^{*} N
\rightarrow \Delta$ reaction. In the next two years the question
transformed into a  more precise determination of these effects.

The idea of organizing these two events grew out of a series of
informal "OOPS workshops", small gatherings of a dozen or so
theorists and experimentalists which occurred during the summers
of 2001 and 2002  at MIT and which were focused on addressing the
physics pursued by the Out Of Plane Spectrometer (OOPS) of Bates.
It was realized that the field had reached a stage of maturity,
where the investigation of the structure of the proton, and of
hadrons in general, demanded that old questions be viewed in the
light of new discoveries in related fields. New, more powerful
experimental arrangements were becoming operational employing
beams of superb quality and polarization. The impressive data
emerging at the high end of the spectrum, at a multi GeV scale,
had a bearing on the same questions but we lacked the theoretical
framework to connect them. The precision of the experimental
data, at least in certain sectors, had reached the stage where
their interpretation was the limiting factor. Phenomenological
modeling had reached a level of sophistication which allowed the
distillation of key physical parameters out of a massive and
diverse set of data.  Theory had, for the first time, either
through Lattice QCD calculations or through Effective Field
Theoretical approaches, made contact to experiment. These
developments needed to be understood as pieces of the same
puzzle, that of the structure of hadrons. The issue of "shape" as
an appealing unifying theme.

Many issues emerged clearly in the first meeting at MIT: to define
precisely the meaning of "shape" for a hadron and to identify the
observables that can access information concerning it; the need to
understand the role of the pion cloud and of chirality (and its
spontaneous breaking); to quantify the degree to which the color
magnetic interaction amongst quarks can explain deformation; to
understand the role of relativity;  to provide the connection
amongst physical quantities (such as elastic form factors,
transition form factors, and polarizabilities) in determining the
shape of hadrons and explaining the mechanisms that determine it.
Finally, to address the elusive connection between phenomena which
have driven the field at high energies (such as the EMC effect and
the "spin crisis" ) and the low energy investigations.

In this volume these issues are raised and addressed. In this
overview we attempt to provide both an introduction and a road
map of how to read this volume. It is meant to be accessible, and
to a certain degree, pedagogical for young scientists entering the
field.

\section{Introduction\label{sec:intro}}

The possibility that hadrons would have non-spherical amplitudes
was first suggested by Glashow in 1979 on the basis of non-central
(tensor) interactions between quarks \cite{glashow}. This
conjecture was based on the premise that there is a color
spin-spin interaction between the quarks\cite{Ru75}  which is
modeled after the interaction between magnetic dipoles in
electromagnetism ("Fermi-Breit" interaction) \cite{hyperfine}.  A
few years later Isgur, Karl, and Koniuk wrote a paper "D Waves in
the Nucleon: A Test of Color Magnetism" \cite{isgur1} which
detailed an impressive list of empirical evidence for this
hypothesis. In this paper they singled out the E2 transition in
the  $\Delta \rightarrow N \gamma$ transition as being the most
definitive test of this hypothesis.  Of additional interest are
the quark model calculations  which showed that the D state
admixtures caused by the color hyperfine interaction predict a
non-zero neutron charge distribution and  RMS charge
radius\cite{rn,isgur2}. These theoretical speculations induced
concerted experimental and theoretical efforts to measure and
calculate deviations from spherical symmetry (non-spherical
amplitudes) in hadrons.

Since the proton has spin 1/2, one cannot observe a static
quadrupole moment, and this has made the effort much more
difficult. Since the $\Delta$ has spin 3/2,  the
$\gamma^*N\rightarrow \Delta$ reaction has been actively studied
for non-zero  quadrupole amplitudes in the nucleon and $\Delta$.
Due to spin and parity conservation in the
$\gamma^*N(J^\pi=1/2^+) \rightarrow \Delta(J^\pi=3/2^+)$ reaction,
only three multipoles can contribute to the transition: the
dominant magnetic dipole ($M1$), the electric quadrupole ($E2$),
and the Coulomb quadrupole ($C2$) photon absorption multipoles.

Following these initial calculations of non-spherical hadron
amplitudes\cite{isgur1,rn,isgur2} there has been a considerable
experimental and theoretical effort to study quadrupole
transitions in the $\gamma^{*} N \rightarrow \Delta$ reaction
(for reviews, in addition to the present volume, see
\cite{nstar2001,cnp,amb,bl, pvy}).  Experiments have been
performed with real <Kotula>\footnote{All references to articles
in this volume will be shown with a <bracket>. For  contributions
with more then two authors only the first name will be
cited.}\cite{beck,blanpied} and virtual photons <Ungaro,
Sparveris, Smith> \cite{warren}- \cite{Ungaro}.
In addition many calculations have been performed including
predictions of hadron deformation <Vanderhaeghen> and reaction
models which enable us to extract the resonance amplitudes from
the experimental data <Drechsel-Tiator>.

In the quark model, the non-spherical  D state amplitudes in the
nucleon and $\Delta$ have a probability of $\simeq$ 1\%
\cite{isgur1}. The  predicted E2 and C2 amplitudes using quark
models\cite{pvy, capstick_karl} are much too small to explain the
experimental results  and even the dominant M1 matrix element is
$\simeq$ 30\% low <Giannini>. A likely cause of these dynamical
shortcomings is that the quark model does not respect chiral
symmetry, whose spontaneous breaking leads to strong emission of
virtual pions (Nambu-Goldstone Bosons)\cite{amb}. These couple
to  nucleons as $\vec{\sigma}\cdot \vec{p}$ where  $\vec{\sigma}$
is the nucleon spin, and $\vec{p}$ is the pion momentum. This
coupling is strong in the p wave and mixes in non-zero angular
momentum components.  The emission and absorption of pions from
the nucleon leads to the well known long range N-N tensor
interaction. It was first pointed out in 1983 that the addition
of virtual pions would significantly enhance the magnitude of the
quadrupole amplitudes\cite{Eisenberg}. Following that, there were
a number of calculations in which virtual pions were added to the
quark contribution (for a review see\cite{pvy}). More recently it
has been shown that the shortfall of the quark models for the
$\gamma^{*} N \rightarrow \Delta$ transition can be compensated
for by the  effects of the pion in dynamic models
<Drechsel>\cite{sato_lee,dmt} or in chiral effective field theory
calculations <Gail-Hemmert, Pascalutsa-Vanderhaeghen>\cite{pasc,
gh}. Nucleon charge distributions are also altered by the
emission and absorption of virtual  pions. For the neutron the
$\pi^{-}p$ virtual intermediate state gives rise to the observed
negative square RMS charge radius, $r_{n}^{2}$,  since the pion
tends to be at larger radii than the proton.

\begin{figure}[htb]
\includegraphics[width=1.00\textwidth,height=0.74\textwidth,angle=-00]{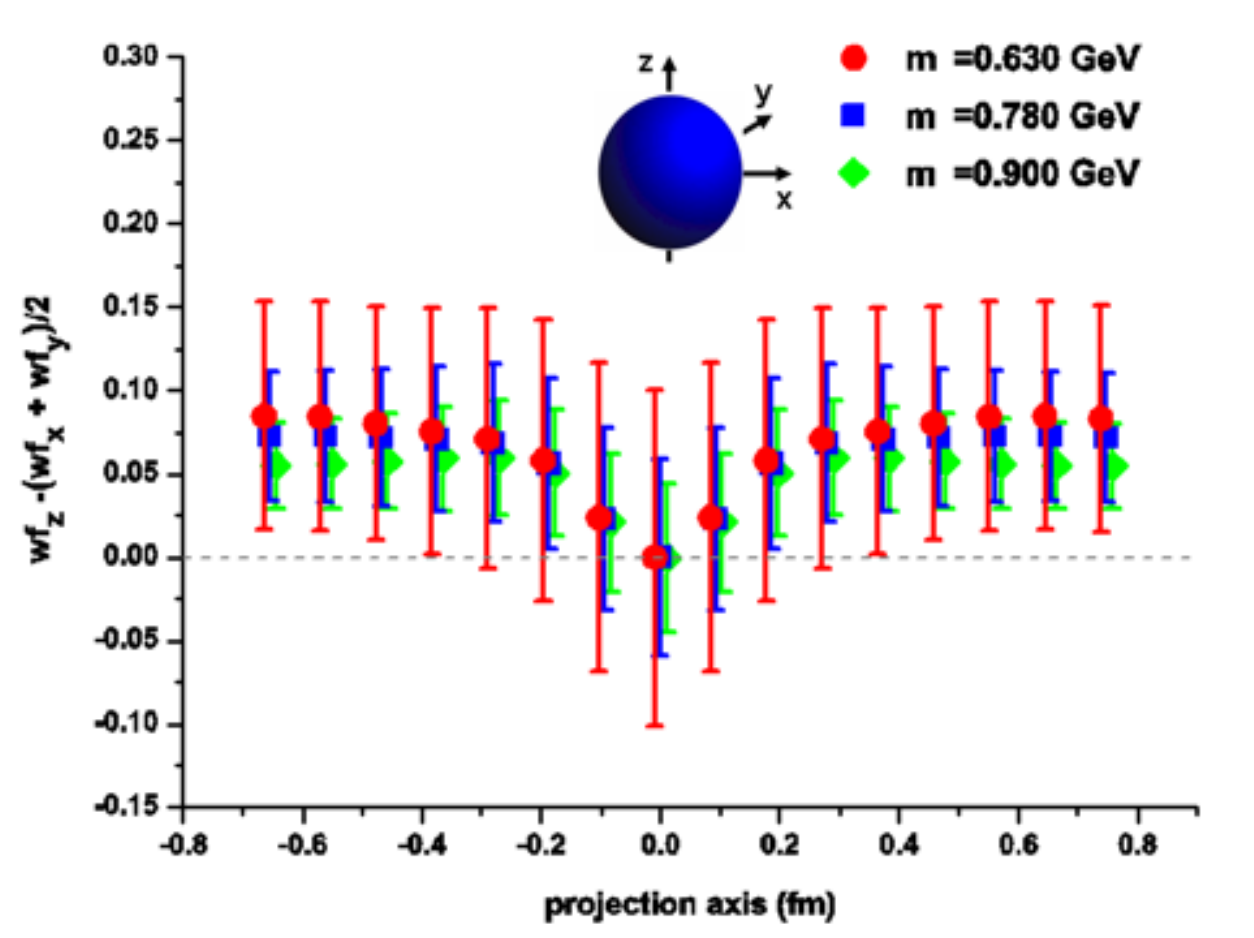}
\caption{Lattice calculation of the deformation of the $\rho$
meson<Alexandrou>\cite{Dina}. The plot shows the fractional
difference of the expectation value of the density along z ( the
spin direction) compared to the expectation in the perpendicular
direction versus distance along the z axis. The calculation has
been performed for three values of $m_{\pi}$. }
\label{fig_rho_lattice}
\end{figure}

Of general interest,  lattice calculations<Alexandrou>
\cite{Dina} show that hadrons, such as the $\Delta$ and the $\rho$ meson are deformed. In Fig.~\ref{fig_rho_lattice} it is shown that the
wavefunction of the $\rho$ meson has a prolate shape, i.e. it
extends further along the spin axis. This trend increases as the
pion masses employed become lighter. For the $\Delta$ an oblate
shape has been found<Alexandrou>  \cite{Dina}. Lattice QCD
calculations have also been performed for nucleon form factors
and for the $\gamma^{*} N \rightarrow \Delta$ transition
<Alexandrou>\cite{Dina}. These have been performed with large
quark masses which are beyond the convergence of  chiral
extrapolations. The present state of the art is that, within the
lattice and chiral extrapolation errors, there is rough agreement
with experiment. Clearly this is an area where further work is
required.

In a related field, the angular momentum content of the proton
measured in deep inelastic scattering has been the subject of
intense experimental and theoretical activity since it was
discovered that only $\simeq$ 25\%  of the spin comes from the
valence quarks\cite{spin-review}. There are concerted efforts to
measure the spin contribution of the $\bar{q}q$ quark sea, the
gluons and the angular momentum of the quarks. Recently, the
Sivers  distribution function has been measured by the HERMES
collaboration and found to be non-zero <Rith> . This  describes
the correlation between intrinsic quark $p_{T}$ and transverse
nucleon spin and the result requires L > 0 orbital angular
momentum components  in the nucleon <Rith>.

There have been developments in mapping out the shape of the
proton through Generalized Parton Distributions (GPD), which can
be measured in deep  virtual Compton scattering and hard
exclusive meson production <Vanderhaeghen>. A Fourier transform
of the the GPD yields a tomographic view of the nucleon, i.e. the
simultaneous distribution of quarks with longitudinal momentum x
and transverse position b <Kroll>. Sum rules (integrals)  of the
GPD's are the elastic scattering form factors\cite{GPD-FF}. It is
also possible to define an $N \rightarrow \Delta$ GPD and in a
similar fashion one can obtain the $\gamma^{*} N \rightarrow
\Delta$ form factors from the appropriate integrals. These  are
in reasonable agreement with experiment <Vanderhaeghen>.

\subsection{Shape of Hadrons}

It is worthwhile digressing to discuss the issue of shape of
hadrons which is often misunderstood. The reason may be rooted in
the fact that in quantum field theory, unlike non-relativistic
quantum mechanics, form factors are not the Fourier transform of
static densities. This is due to relativistic effects and to
quantum fluctuations. At distances of the Compton wavelength
($\simeq$ 0.2 fm for the nucleon) one expects classical pictures
to break down. Strictly speaking, one can discuss form factors
and moments of ground-state and transition amplitudes. As an
example,  the concept of size is straightforward  since it is
generally understood to mean the RMS radius of a hadron.
Nevertheless, shape is a useful physical concept and can be
precisely quantified in terms of ground- state or transition
moments. Non-spherical distributions can be measured with the
observation of quadrupole amplitudes.

The notion of shape is sometimes discussed in analogy to nuclear
physics where deformed nuclei with rotational bands are well
known. Somewhat closer to the present consideration would be spin
1/2 nuclei with vibrational and rotational degrees of freedom; as
in the case of the proton, their static quadrupole moment of the
ground state is zero. Their shape is derived through the study of
their excitation spectrum. For hadrons like the proton and
$\Delta$ the situation is far more complex since the constituents
fluctuate (e.g. there are virtual $\bar{q}q$ pairs or mesons),
relativity is important, and the system cannot be approximated as
a rigid rotor. For the proton the static quadrupole moment
vanishes identically on account of its spin, even if there is a
probability of D states in the quark model and p states for
virtual pion emission.  This is an important point since we
require any theory that describes these hadrons to predict these
non-spherical magnitudes. The difficulty lies in the testing of
these predicitions because we need to study them through
transitions to well described excited states. In the spectrum of
the nucleon the only isolated state is the $\Delta^{+}(1232)$,
which of course explains the intense experimental interest in the
study of the  $\gamma^{*}N \rightarrow \Delta$ reaction.
Fortunately the transition from J = 1/2 to J = 3/2 with no change
in parity also allows us to observe quadrupole E2 and C2
transition moments

For hadrons with total angular momentum $J \geq 1$ it is possible
to measure a static quadrupole moment Q, which, if non-zero, is a
clear indication of a deviation from spherical symmetry. A
beautiful example of this is shown in Fig.~\ref{fig_rho_lattice}
and in the figures in the contribution in this volume of
Alexandrou, which shows lattice calculations for the $\rho$ meson
($J= 1$) which is prolate (Q > 0, cigar like) and the $\Delta (J=
3/2)$ which is oblate(Q < 0, pancake like).

We can obtain an estimate of the quadrupole moment of the
$\Delta$ using large $N_{c}$ relations. In this case the
relationship between the static quadrupole moment
$Q_{\Delta^{+}}$ and the transition quadrupole moment $Q_ {\gamma
p \rightarrow  \Delta^{+}}$ is given by~\cite{Buchmann-Nc}:
\begin{eqnarray} Q_{\Delta^+} \,\simeq \, \frac{2 \sqrt{2}}{5} Q_{p
\to \Delta^+} = - ( 0.048 \pm 0.002  ) \; \mathrm{fm}^{2},
\label{eq:q-delta}
\end{eqnarray} where the  value of the transition quadrupole moment $Q_
{\gamma p \rightarrow  \Delta^{+}}$ was obtained from the
empirical value of the E2 amplitude\cite{pvy,PDG}. This  formula
should be accurate up to corrections of order $1/N_c^{2}$. The
negative value of $Q_{\Delta^{+}}$ supports the lattice
calculation of an oblate distribution <Alexandrou>.

Since we have found that the pionic degrees of freedom are
important to understand the $\gamma^{*} N \rightarrow \Delta$
reaction, it is instructive to discuss the shape of the nucleon
and $\Delta$ in terms of an old  model which allows the nucleon
to emit and absorb virtual pions. 
This originated  about 1950 to
explain such phenomena as the anomalous magnetic magnetic moments
of nucleons and was called the "pion atomic model"\cite{Feld}.
Although this model has some of the basic physics, it is only the
first term in an expansion of the wave function which
realistically includes multiple pions, other mesons, $N^{*}$
excited states, etc. More recently the pion-cloud model has been
utilized to discuss the issue of hadron shape in the context of
the $\gamma N \rightarrow \Delta$ reaction\cite{BH}. In this
model we describe the nucleon with spin up and the $\Delta$ (with
J =3/2, M=1/2) as:
\begin{eqnarray}
\label{pionwave} \vert N \uparrow \rangle &= & \alpha  \vert
N_{core} \uparrow \rangle  + \beta( -\sqrt{1/3} \vert N_{core}
\uparrow R(r_{\pi})Y1_0(\Omega_{\pi}) \rangle  +\sqrt{2/3} \vert
N_{core} \downarrow R(r_{\pi})Y1_1(\Omega_{\pi})  \rangle ),
\nonumber \\ \vert \Delta^+ \uparrow \rangle &= & \alpha'  \vert
\Delta_{core}^{+} \uparrow \rangle  + \beta' (\sqrt{2/3} \vert
N_{core} \uparrow R(r_{\pi})Y1_0(\Omega_{\pi}) \rangle
-\sqrt{1/3} \vert N_{core} \downarrow
R(r_{\pi})Y1_1(\Omega_{\pi})  \rangle ),
\end{eqnarray} where the nucleon and $\Delta$ cores can be
approximated as s-wave quarks. In this case the quadrupole
transitions in the photo- and electro $\Delta$ excitation come
from the charged pion components of the $\vert N_{core} \pi
\rangle$ wave function which are obtained  using the proper
isospin Clebsh-Gordon coefficients \cite{BH}. When this is done,
the   spectroscopic $\Delta$ and $\gamma N \rightarrow \Delta$
quadrupole moments are found to be\cite{BH}
\begin{equation}
\label{pionquad} \label{pcm1} Q_{\Delta^+}  = -{2 \over 15} \,
{\beta'}^{2}\, r_{\pi}^{2}, \qquad Q_{\gamma +p \rightarrow
\Delta^+}  = {4 \over 15} \, {\beta'} \beta \, r_{\pi}^{2},
\end{equation} where $r_{\pi}2$ is the RMS radius of the pion
distribution $R(r_{\pi})$.  It is interesting to see that
$Q_{\Delta^+}$ < 0,  independent of the sign of $\beta^{'}$, and
in agreement with Eq.~\ref{eq:q-delta}. It is also interesting to
see that to have $Q_{\gamma +p \rightarrow \Delta^+} \neq $ 0
both $\beta$ and $\beta^{'}$  must be non-zero, i.e. the
non-spherical pion-cloud must occur both in the nucleon and the
$\Delta$.

From Eq.~\ref{pionwave} the pion density of the proton $\rho
(r_{\pi})$ is:
\begin{eqnarray}
\label{piondensity} \rho (r_{\pi}) = \beta^{2} R(r_{\pi})^{2} (
1/3 \vert Y1_0(\Omega_{\pi}) \vert ^{2} +  2/3 \vert
Y1_1(\Omega_{\pi})\vert^{2}  )  \propto  \beta^{2} R(r_{\pi})^{2}(
\sin(\theta_{\pi})^{2} + \cos(\theta_{\pi})^{2}).
\end{eqnarray} From this simple example, we see that although there are
non-spherical amplitudes in the proton, proportional to $\beta$,
on average it must be spherical due to its J = 1/2 character. As
can be seen from Eq.~\ref{pionquad} the non-spherical amplitude
is observable in the quadrupole part of the $\gamma N \rightarrow
\Delta$ transition.

 Buchmann and Henley, in analogy to the non-
relativistic case of deformed nuclei have proposed "undoing" the
Clebsh-Gordon coefficients in Eq.~\ref{piondensity} to obtain a
non-spherical intrinsic pion density\cite{BH}.  Once different
values are substituted one can obtain either oblate or prolate
shapes. Their choice gives
the proton  a prolate shape in its "intrinsic frame". Based on
Eq.~\ref{pionquad} the $\Delta$ is oblate. For a simple geometric
picture it is probably preferable to think about the time
dependence of the pion-cloud where it oscillates from the prolate
$ Y1_0(\Omega_{\pi})$ to the oblate $ Y1_1(\Omega_{\pi})$ shape
as in Eq.~\ref{piondensity}

Miller has proposed that the angular momentum content of the
nucleon can be seen  in a (non-relativistic) spin dependent
density $\rho(r,\vec{n})$\cite{Miller,Miller2}
\begin{eqnarray}
\label{pol-density} \rho(r,\vec{n})  =   \rho(r)[1+
\vec{\sigma}\cdot \vec{n}]/2,
\end{eqnarray}
where $\vec{n}$ is the direction of the parton spin and
$\vec{\sigma}$ the nucleon spin. The relativistic version of this
has also been presented\cite{Miller, Miller2}. This
(non-relativistic) spin dependent density can be evaluated  for
the pion-cloud model of Eq.~\ref{pionwave} for  the nucleon spin
in the virtual pion-nucleon state $\vec{n}$ being parallel or
anti-parallel to the nucleon spin.
\begin{eqnarray}
\label{pol-density-pion-cloud} \rho(r,\vec{n}= \vec{s})
\propto   \beta^{2}
R(r_{\pi})^{2} \sin(\theta_{\pi})^{2}, \nonumber \\
\rho(r,\vec{n}= -\vec{s})   \propto   \beta^{2} R(r_{\pi})^{2}
\cos(\theta_{\pi})^{2}.
\end{eqnarray}
It can be seen that the angular momentum of the $\pi$N state
shows up very clearly in $\rho(r,\vec{n})$. The integral of the
relativistic version of the spin-dependent density can be related
to the results of spin dependent measurements in deep inelastic
scattering $\Delta u, \Delta d, \Delta s$\cite{Miller, Miller2}.
The problem (challenge) is that  no experiment has been
conceptualized to exhibit this density. As can be seen from the
pion-cloud model, and from calculations on nucleon densities
using models that fit the elastic e p scattering, there are
potentially interesting spin dependent densities that can be
measured\cite{Miller,Miller2}.

Last, but not least, we mention  the effects of relativity on
shape. As is well known, in special relativity the shapes of
moving objects change when they move at high speeds. If one
considers the hydrogen atom, or positronium, using the Dirac
equation the small component of the wave function is in a p wave.
This leads to a small, but definite, non-spherical
amplitude\cite{Miller,Miller2}. The issue of relativity for
positronium is discussed in more detail using the Bethe-Salpeter
equation <Hoyer>. It is shown that the Fock components have a
more complex shape change under relativistic boosts than a
flattening of their distributions along the direction of the
boost. To our knowledge this effect has not been incorporated
into quark models with current quark masses for which the
relativistic aspects should be large.

\section{Theory and  Models of Non-Spherical Hadron Amplitudes}
In the introduction the development of the field and the key ideas
and calculations were presented in historical order. In this
section the basis for non-spherical hadron amplitudes will be
presented in somewhat more depth, primarily from the point of
view of QCD. An important theoretical development of the past few
years is the emergence of lattice simulations of QCD and their
application to nucleon form factors and to the $\gamma^{*} N
\rightarrow \Delta$ reaction <Alexandrou>\cite{Dina}. At the
present time these calculations are limited to relatively heavy
quark masses by the constraints of computational ability
(computer resources and technology). The calculations reported
here correspond to pion masses in the range of 410 to 560 MeV  for
quenched and 380 to 690 MeV for unquenched calculations
<Alexandrou>\cite{Dina}. These masses are too high to see the
pion-cloud effects and also for reliable chiral extrapolations.
Unfortunately, the time required to improve these calculations
goes as a high inverse  power of the  pion mass. Computational
limitations also  have resulted in relatively large statistical
errors encountered in the sampling of the numbers of gauge
configurations ($N_{g}$) in these calculations with the errors
going as $1/\sqrt{N_{g}}$.  In addition, the effects of employing
different lattice Fermion formulations and lattice spacings must
be evaluated <Alexandrou>\cite{Dina}. Using chiral
extrapolations, qualitative agreement with experiment has been
achieved <Pascalutsa>\cite{pasc}.  These calculations predict
significant deviations of hadron structure from spherical
symmetry  (shown in Fig.~\ref{fig_rho_lattice}) and in the
non-zero values of the quadrupole amplitudes in the $\gamma^{*} N
\rightarrow \Delta$ reaction. Predictions for the axial form
factors in the parity violating electro-excitation of the
$\Delta$ have also been made in advance of the experiments, which
are to be performed at JLab.  It is gratifying that calculations
with pion masses as low as 250 MeV are becoming available
<Alexandrou>. These will enable accurate chiral extrapolations to
connect with data in a quantitative way and also allow us to
better explore the pion contribution to the observables.

     In the past few years theoretical calculations have progressed
using chiral effective field theory <Gail-Hemmert,
Pascalutsa-Vanderhaeghen>\cite{GHKP,gh,pasc,pvy}. These
calculations represent the first approach to estimate QCD
predictions using a perturbation expansion. The first to be
published used the epsilon expansion, where   $\epsilon =
\{\frac{|Q|}{\Lambda},\frac{m_{\pi}}{\Lambda},\frac{\Delta}{\Lambda}\}$
with      $\Lambda = \{4\pi F_{\pi}, M_{N} \} \simeq $
1GeV\cite{GHKP,gh}. These calculations of the (complex) M1, E2,
and C2 transition amplitudes were carried out to order
$\epsilon^{3}$ and are in qualitative agreement with experiment
as will be shown below. An alternative expansion in terms of
another small parameter $\delta$ was also carried out
<Pascalutsa-Vanderhaeghen>\cite{pasc,pvy}. In this case  $\delta
= \{\frac{|Q|}{\Lambda},\frac{\Delta}{\Lambda}\}, \delta^{2}=
\frac{m_{\pi}}{\Lambda}$, where  $\Lambda= 4\pi F_{\pi}$. The
main difference between these calculations is that in the delta
expansion $m_{\pi} \simeq$ 140 MeV is considered significantly
smaller than $\Delta  \simeq$ 300 MeV. In practice this means
that some graphs are neglected as higher order in the $\delta$
expansion. From a theoretical perspective a significant
difference between $m_{\pi}$ and $\Delta$ is that only  the
former vanishes in the chiral limit,  and it is argued that it
should be treated on a different footing
<Pascalutsa-Vanderhaeghen>\cite{pasc,pvy}.  (Note that $\Delta$
vanishes in the large $N_{c}$ limit.)   The other difference  is
that the $\epsilon$ expansion has also been employed in heavy
Baryon chiral effective field theory, while the $\delta$
expansion has been utilized with a relativistic version of chiral
perturbation theory. Furthermore, the way that the low energy
constants have been chosen is different in both calculations. It
is also the case that the calculations using the $\delta$ scheme
have been carried out successfully for the cross sections  in the
$\gamma^{*} N \rightarrow \Delta$ reaction. Within the estimated
relatively large errors due to the neglect of higher order terms,
both approaches agree with the data for the observables that  have
been calculated. To make further progress we need to have the
next order calculations (including cross sections) for both
expansions. We  can then use the comparison with experiment to
see if there are any significant differences between these two
approaches. This is particularly important since the differences
between the two calculations involve a mixture of ingredients
which makes them somewhat difficult to compare purely
theoretically.

\begin{figure}
\begin{minipage}{3.5cm}
\hspace*{-6.1cm}
\includegraphics[height=.27\textheight]{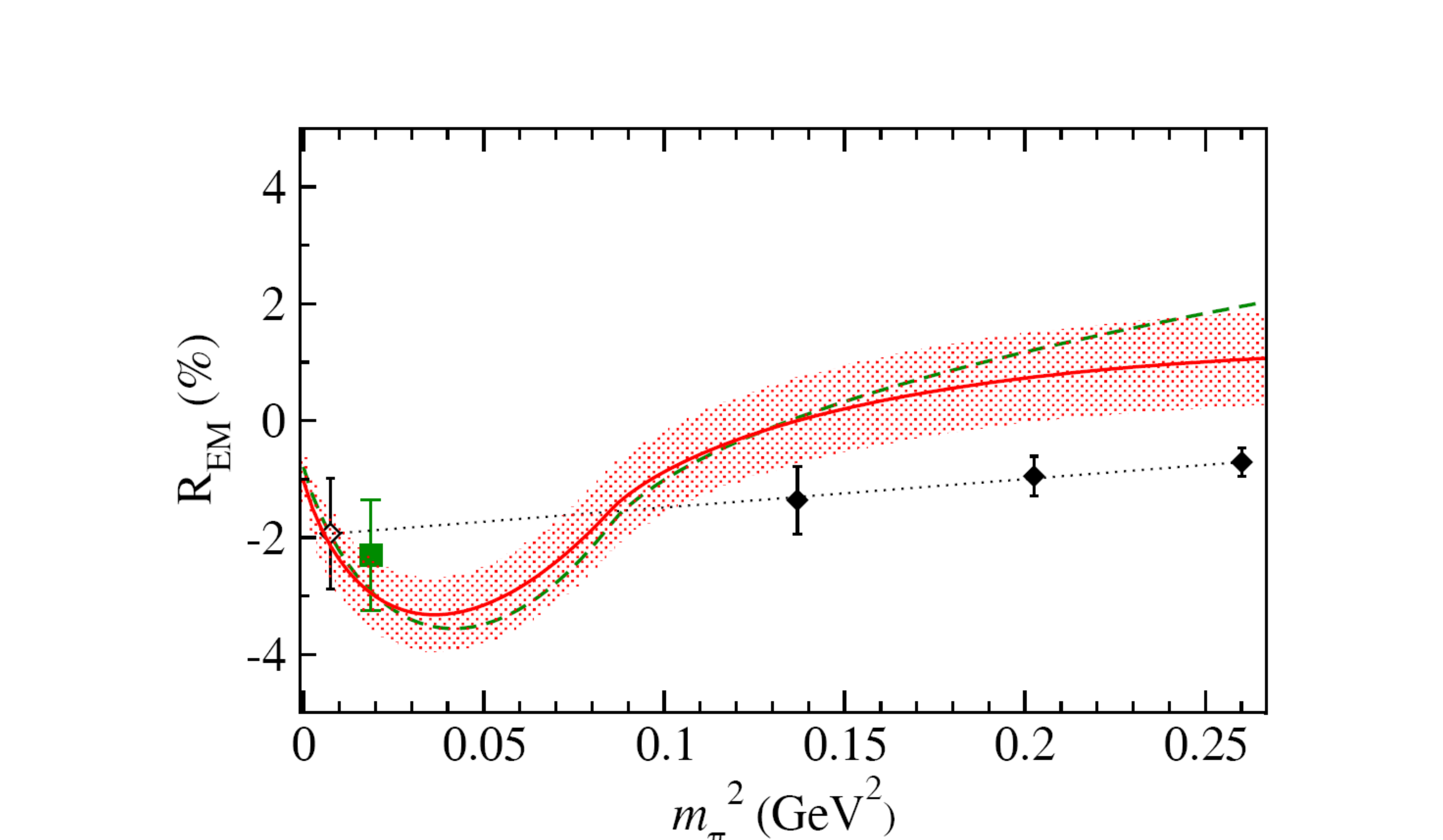}
\end{minipage}
\begin{minipage}{3.5cm}
\vspace*{0.90cm}
\includegraphics[height=.24\textheight]{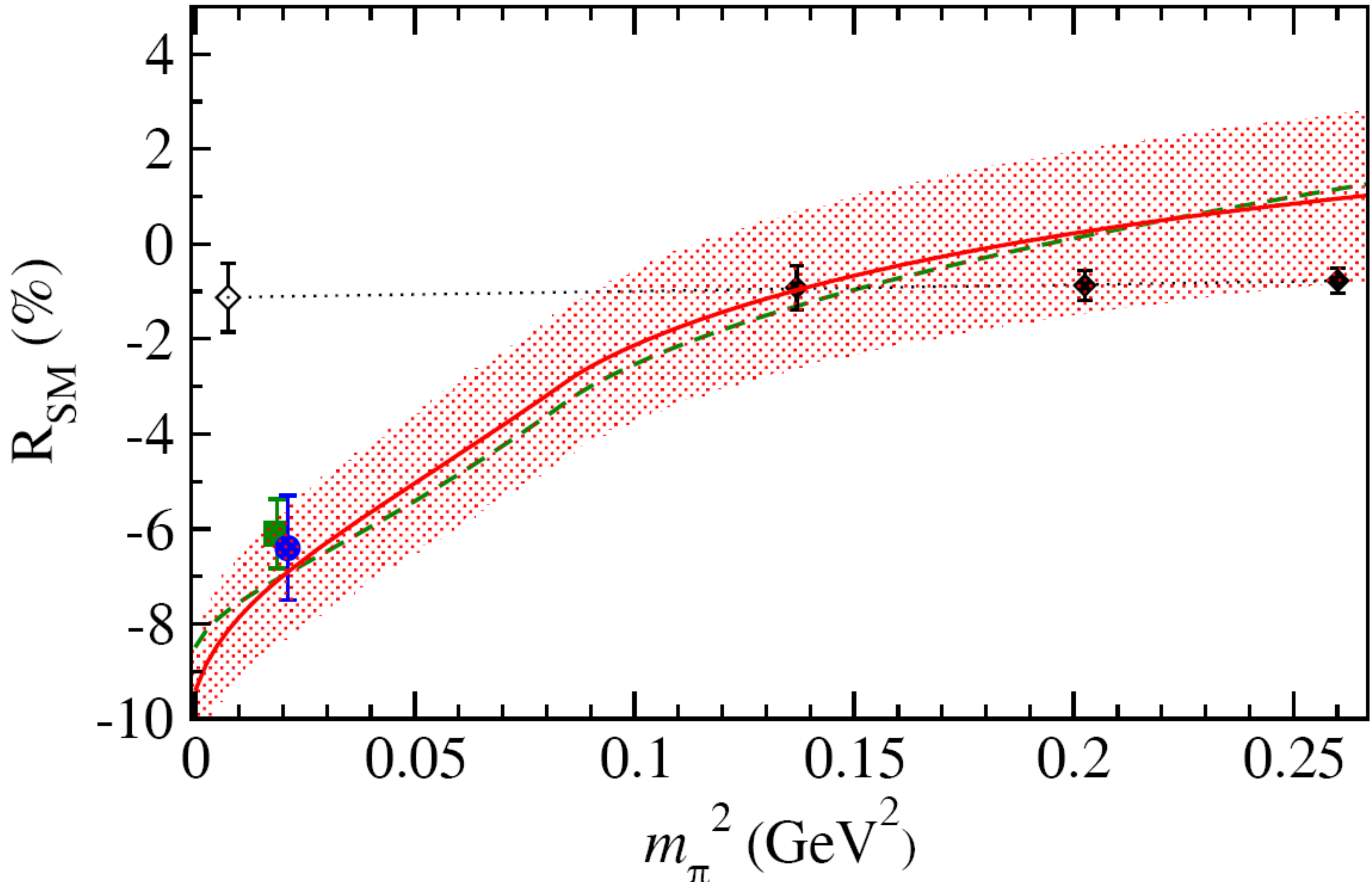}
\end{minipage}
\vspace*{0.5cm} \caption{The pion mass dependence of $R_{EM}$
(left panel) and $R_{SM}$ (right panel), at $Q^2=0.1$ GeV$^2$. The
blue circle is a data point from MAMI\cite{pospischil}, the green
squares are data points from BATES \cite{mertz,sparveris}. The
three filled black diamonds at larger $m_\pi$ are lattice
calculations~\protect\cite{Dina}, whereas the open diamond near
$m_\pi \simeq 0$ represents their extrapolation assuming linear
dependence in $m_\pi2$. Red solid curves: NLO result when
accounting for the $m_\pi$ dependence in $M_N$ and $M_\Delta$;
green dashed curves: NLO result of <Pascalutsa-Vanderhaeghen>
Ref.~\cite{pasc}, where the $m_\pi$-dependence of $M_N$ and
$M_\Delta$ was not accounted for. The error bands represent the
estimate of theoretical uncertainty for the NLO calculation. See
text for discussion.} \label{fig:REM-chiral}
\end{figure}

The $\delta$ scheme has been used to predict a chiral
extrapolation for the lattice calculations as a function of pion
mass as shown in Fig.\ref{fig:REM-chiral}. This has been done for
$m_{\pi}$ values from 0 up to 0.35 - 0.5 GeV where the lattice
calculations were performed. Even though these values far exceed
the premise of the chiral calculations that $m_{\pi} << \Delta
\simeq$ 0.3 GeV they produce qualitatively similar results to the
calculated lattice values for the EMR= E2/M1 and CMR= C2/M1
ratios. The calculated curves show a strong dependence for
$m_{\pi} \leq $ 0.35 GeV. They also indicate  that a linear dependence
in $m_{\pi}^{2}$ may be accurate for the EMR but not the CMR.
Using this chiral
extrapolation<Pascalutsa-Vanderhaeghen>\cite{pvy, pasc}  improves
the agreement between the lattice calculations and experiment.
These first results, impressive as they are, they can only be
viewed as suggestive due the above mentioned limitations. They
pave the way and they highlight for the need for lattice
calculations with smaller pion masses and, in addition, higher
order chiral calculations to make quantitative comparisons
between QCD and experimental data.

As was discussed in the introduction, historically the first
calculations of the deviation of hadron shapes from spherical
symmetry were performed using the non-relativistic quark
model\cite{isgur1, rn, isgur2}. Although the quark model is able
to reproduce the spectrum of low lying excitations of hadrons, it
is not able to accurately reproduce the observed transition
rates\cite{capstick_karl}. In addition to the problems of not
agreeing with experiment, the quark model has internal
consistency problems. It has been shown that depending in the
choice of  gauge the size  of the calculated E2 transition matrix
element can differ by an order of magnitude \cite{gauge}. This
problem, which is a manifestation of space truncation, is
particularly serious when only 2$\hbar \omega$ configurations are
used. Subsequent quark model calculations using up to 6$\hbar
\omega$ configurations were performed and showed less sensitivity
to the choice of charge and current
operators\cite{capstick_karl}. These calculations produced EMR
ratios of $\simeq$ -0.4\% at the photon point for the
non-relativistic quark model and $\simeq$ -0.2\% for the
relativized quark model\cite{capstick_karl} as compared to the
experimental value of $\simeq$ -2.5\%. Since the value of the M1
transition is typically underestimated  by $\simeq$ 30\% in quark
models <Giannini>\cite{pvy, capstick_karl}, this makes the
underestimate of the E2 matrix element even more serious. As will
be discussed in the next section, the $Q^{2}$ dependence of the
quark model calculations are even in more disagreement with
experiment. This holds for all of the non-relativistic and
partially relativized quark models.

In being able to describe  low lying energy levels but not
electromagnetic transition rates, the quark model is similar  to
the shell model of nuclei which  can also reproduce the low lying
spectra of excited states but requires collective effects to
explain the electromagnetic transition rates. In the case of
hadrons, however, the solution to the problem is quite different.
The quark model does not respect the spontaneously hidden chiral
symmetry of QCD which leads to the p wave emission and absorption
of pions. This non-spherical $\pi$N interaction leads to greatly
enhanced quadrupole transitions in the $\gamma^{*} N \rightarrow
\Delta$ reaction and also to the proper enhancement of the M1
transition\cite{sato_lee,dmt}.

The fact that pions were needed to cure the dynamics problems of
the quark model was first suggested in 1983 in the context of a
cloudy bag model\cite{Eisenberg}. Subsequent calculations with
chiral (cloudy) bags gave E2/M1 values of $\simeq$ -2.0\% at the
photon point but with the M1 matrix element still $\simeq$ 30\%
low\cite{cloudy2}. These predictions are unstable as demonstrated
by the fact that  calculations with a similar model, but taking
center of mass recoil into account, obtained an M1 matrix element
close to experiment but reduced E2/M1 to $\simeq$
0\cite{cloudy3}. The pion-cloud was also shown to be able to
produce E2/M1 ratios $\simeq$ -2\% at the photon point with a
linear sigma model, but the magnitude of the M1 transition is
$\simeq$ 20\% low\cite{Sirca} and the $Q^{2}$ dependence does not
match experiment (for a  review of the quark models and pionic
extensions see \cite{pvy}).

An extension of the quark model to include $\pi$ and $\sigma$
meson exchanges has been developed and leads to a prediction of
E2/M1 $\simeq$ -3.5\%  at the photon point\cite{Buchmann}. The
enhancement of the E2 amplitude comes from a two quark operator
in this model. However, the magnitude of the M1 amplitude is
$\simeq$ 30\% low, a  value which is not improved over other
non-relativistic quark models. Within this model an interesting
relationship has been derived between $G_{En}(Q^{2})$, the
electric form factor of the neutron, and the $C2(Q^{2})$
transition amplitude which is reasonably accurate empirically
<Buchmann>\cite{Buchmann2}; in the low $Q^{2}$ region  the
transition quadrupole moment of the $\gamma p \rightarrow
\Delta^{+}$ reaction is related to the square of the RMS radius
of the neutron charge distribution $r_{n}^{2}$ as:
\begin{eqnarray} Q_{\gamma p \rightarrow \Delta} = r_{n}^{2}/\sqrt{2}
=-(0.084 \pm 0.002)~{\rm fm}^{2},  \nonumber \\ Q_{\gamma p
\rightarrow \Delta}^{exp} = -(0.085 \pm 0.003) ~{\rm fm}^{2},
\label{eq:QND}
\end{eqnarray}
\noindent  where the value of $r_{n}^{2}= -0.119 \pm 0.003~{\rm
fm}^{2}$ was taken from \cite{Kees-review} and the experimental
value of $Q_{\gamma p \rightarrow \Delta}$ was obtained
\cite{pvy} from the empirical value of the E2
amplitude\cite{PDG}. The remarkably good agreement is impressive.
Equation~\ref{eq:QND} was initially derived using the quark
model\cite{Buchmann}  and subsequently in the large $N_{c}$ limit
<Buchmann>\cite{Buchmann-Nc}.

\subsection{Pion Contribution to Hadron Properties} It is well known
from deep inelastic scattering that the  $\bar{q}q$  amplitudes
play a large role in nucleon structure\cite{spin-review}. On
general grounds it is expected that a significant part of this
contribution will be in the long range contribution of pions.
These are strongly coupled to hadrons, and due to the spontaneous
hiding of chiral symmetry in QCD are emitted and absorbed in p
states. This contributes significantly to the long range non-zero
angular momentum amplitudes and the observables that depend on
them such as the quadrupole transitions in the $\gamma^{*} N
\rightarrow \Delta$ reaction.  However, it is not possible to
quantify the contribution of the pion contribution to any
observable in a model independent way <Meissner>. Qualitatively we
expect  that if the pion contribution to a specific observable is
large, this would show up in a lattice calculation  by a
relatively rapid  dependence  on the pion mass, particularly as
one approaches the physical value.  Again this will be difficult
to quantify using  a chiral extrapolation. In chiral perturbation
theory the pion loop contributions are well defined. However,
there are both short range physics and pionic contributions to
the low energy constants and the relative contribution cannot be
obtained in a model independent way <Meissner>. These constants
are either obtained by fitting chiral calculations to data or to
lattice calculations.   For dispersion calculations it is
straightforward  to identify the single pion contribution but not
possible to isolate the pionic part of the empirical  vector
meson contributions. In the case of quark model calculations
there is no pion contribution. However when empirical form
factors are introduced for the dressed quarks it is not possible
to identify the pionic contribution.

Despite this quantitative limitation, it is still physically
interesting to  qualitatively see where pionic effects are
significant. For example, as has been shown above, in the
quadrupole amplitudes in the $\gamma^{*} N \rightarrow \Delta$
reaction, quark models do not fall far short and models with
pion-clouds are in qualitative agreement with the data. In such a
case we can be reasonably confident that the bulk of the observed
quadrupole amplitudes is due to pionic effects. As will be shown
in the next section, virtual pion effects significantly
contribute in virtual Compton scattering. For nucleon form
factors on the other hand, as  will also be discussed in the next
section, there is also a significant pionic contribution at low
$Q^{2}$. However, it is difficult to exactly identify  the pionic
contribution. The issue of the identifying and quantifying the
role of mesonic, and in particular pionic, contributions is
physically interesting and important. It would be good to see
some progress  in the coming years.

\subsection{Nucleon Form Factors and Virtual Compton Scattering} For
one photon exchange processes nucleon properties cannot be
directly used to demonstrate non-spherical amplitudes due to the
fact that the spin of 1/2 precludes an observable quadrupole
moment. However, low energy processes such as photon and electron
scattering can be used to show pionic effects, which are
responsible for most of the quadrupole amplitudes in the
$\gamma^{*} N \rightarrow \Delta$ reaction at low $Q^{2}$. In
this sense any evidence for pionic effects are part of the shape
of hadrons story.  The importance of pionic effects  can be seen
from chiral perturbation theory (ChPT) calculations for the
nucleon polarizabilities and isovector radii. For both of these
cases these quantities diverge in the chiral limit ($m_{\pi},
m_{u}, m_{d} \rightarrow 0$) indicating that they are pion field
dominated. It also should be pointed out that the long range
tensor part of the Yukawa interaction is due to non-spherical
pion emission and absorption of pions from nucleons.

In the past few years, there has been considerable progress in the
measurement of nucleon electromagnetic form factors <de~Jager>.
There has also been considerable theoretical activity to explain
these results.  Of particular interest to the issue of the pionic
effects on the nucleon form factors is the electric form factor
of the neutron $G_{En}$. Intuitively, it is expected that at low
$Q^{2}$ this is dominated by the virtual emission and absorption
$n \rightarrow \pi^{-} p$. This leads to a negative RMS radius
$r_{n}^{2}$ in agreement with experiment. It is interesting to
note that if one assumes the relationship between the pion-cloud
dominated transition quadrupole moment in the $\gamma p
\rightarrow \Delta$ reaction and $r_{n}^{2}$ presented in
Eq.~\ref{eq:QND}, then this confirms our expectation that the
long range neutron charge distribution is also pion dominated.
This idea is reinforced by quark model calculations which predict
quadrupole transition matrix elements for the $\gamma^{*} N
\rightarrow \Delta$ reaction which are an order of magnitude too
small <Giannini, Sparveris>\cite{sparveris,stave, sparveris2}.
For $G_{En}$ the hypercentral quark model <Giannini> predicts a
maximum magnitude of $\simeq$ 0.15 for $Q^{2} \simeq 0.3~ {\rm
GeV}^{2}$\cite{hcqm-ff} compared to the experimental value of
0.40 to 0.60 <de~Jager>. This significant shortfall implies a
large pionic contribution <Giannini> as in the $\gamma^{*} N
\rightarrow \Delta$ reaction.

\begin{figure}
\includegraphics[height=.5\textheight]{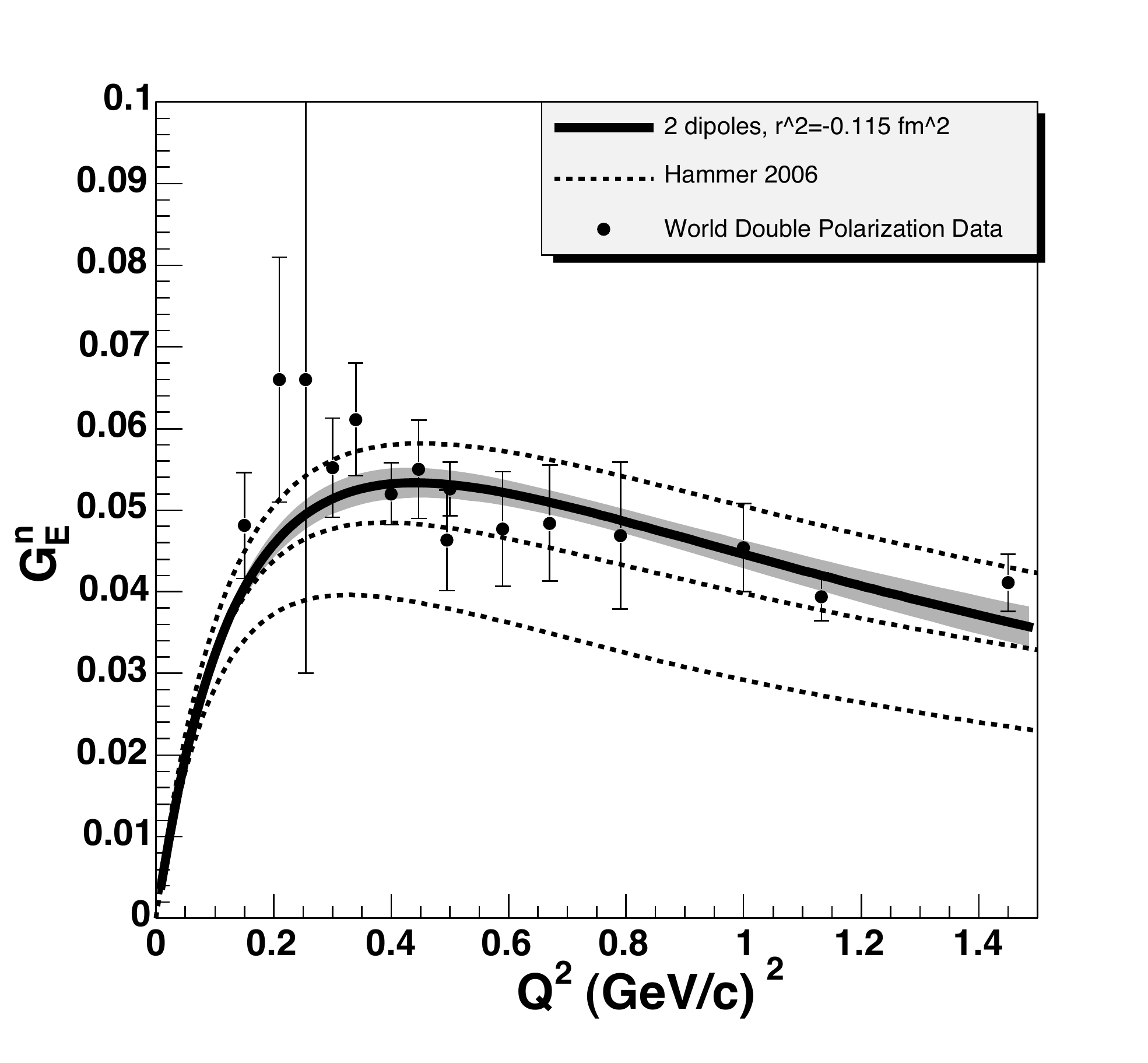}
\hspace*{-1cm}
\includegraphics[height=0.55\textheight]{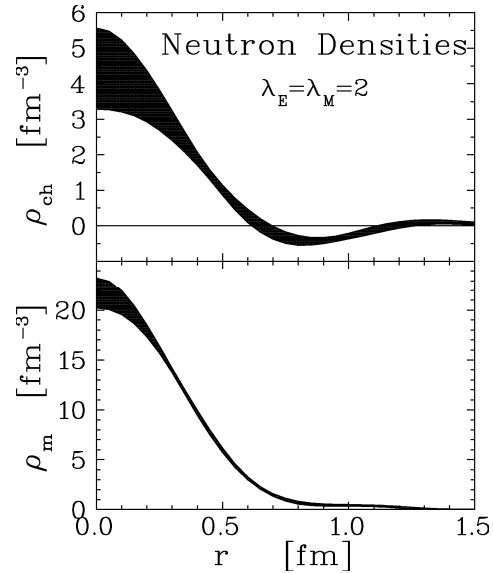}
\vspace*{1cm}
\caption{Left panel: Modern, double polarization data for
$G_{En}(Q^{2})$ (see <deJager> for references). The curves
include the latest dispersion calculation\cite{Hammer,disp} and a
two dipole fit\cite{Kohl} with their one sigma errors. Right
panel: The allowable bands for the neutron  electric and magnetic
densities derived from the observed form factors by
Kelly\cite{Kelly}} \label{fig:GEn}
\end{figure}

The status of $G_{En}(Q^{2})$ is shown in Fig.~\ref{fig:GEn} with
the published polarization data. The curves are the latest
dispersion calculation\cite{Hammer,disp} and  a two dipole
fit\cite{Kohl} with their (not often shown) one sigma errors. It
can be seen that these are in good agreement with the data. The
electric and magnetic nucleon densities  can be obtained, in an
approximate way, from the Fourier transform of the  form factors.
This interpretation is useful physically but is accurate only at
long distances. For high $Q^{2}$ the recoil of the final state
makes the approximation of a static density inaccurate. These
errors are minimized by working in the Breit frame
and the results for the neutron densities~\cite{Kelly} are shown in the right
panel of Fig.\ref{fig:GEn}. For the neutron charge distribution
the fact that the negative charge dominates at long distances, in
agreement with the negative RMS radius squared $r_{n}^{2}$, is in
qualitative agreement with the pion cloud picture in  which $n
\rightarrow \pi^{-} p$ part of the time.

The elastic electron scattering form factors and the $\gamma^{*} N
\rightarrow \Delta$ transition have been calculated in a
relativistic quark model, which includes a pseudoscalar meson
cloud described with chiral techniques, and with additional
vector meson-photon coupling\cite{Faessler, Faessler2}.  The
model includes heavy constituent quarks with empirically
determined dipole form factors (8 parameters) and  current
masses  for the chiral part of the calculation. Perhaps with all
of these parameters it is not surprising that it is in reasonable
agreement with experiment. The calculated pion contribution to
the form factors is found to be significant above  $Q^{2}$ = 0.5
GeV$^{2}$. However, this contribution is artificially
suppressed by a cutoff factor $f_{cut}(Q^{2})$ which is chosen to
go to zero above this value. It should be pointed out that other
model calculations have achieved reasonable agreement with the
data with far fewer ingredients (and parameters). The hypercentral
quark model has achieved reasonable fits to the form factor data
(with the exception of $G_{Ep}$ at low $Q^{2}$) by introducing
empirical form factors for the constituent quarks\cite{hcqm-ff}.
Using vector dominance, reasonable fits to the nucleon form
factors have also been achieved <de~Jager>\cite{Lomon}.
Therefore, it is not clear just what ingredients of the chiral
quark model are required to achieve its good agreement for the
nucleon form factors and the  $\gamma^{*} N \rightarrow \Delta$
reaction for low $Q^{2}$\cite{Faessler,Faessler2}. Further
investigation of this avenue is clearly needed in order to
achieve the most economical description of the data within this
framework.

It is of interest to inquire what information about the shape of
the proton can be inferred from model fits to the nucleon
electromagnetic form factors. Miller has done this with a quark
model fit to the recent form factor data\cite{Miller,Miller2}.
Using the relativistic version of Eq.\ref{pol-density} for the
spin-dependent density the non-spherical shapes of the proton as
a function of quark momenta have been calculated. These
correspond to considerable angular momentum content for the
proton. Recently, there have been some calculations in which  S
wave quark models have been used to fit the nucleon form
factors\cite{Gross}. In a subsequent paper it was pointed out
that this model did have some non-spherical
amplitudes\cite{Miller2}. This inspired a re-fit of the nucleon
form factors with purely S wave quarks\cite{Gross2}.
Unfortunately, due to the spin 1/2 nature of the proton there is
not an unambiguous test of the validity of the angular momentum
content of  these wave functions in elastic ep scattering which
is primarily mediated by one photon exchange. Specifically we do
not have any straightforward way to measure the spin dependent
densities. At this point the most straightforward test would be
for both models to calculate the $\gamma^{*} p \rightarrow
\Delta$ reaction to see if they agree with the data presented
here. Of course, this will also involve a wave-function for the
$\Delta$. Another possibility, using only the proton wave
function, would be to calculate the expected Sivers function for
pion and kaon production <Rith>.

Another reaction which is very sensitive to the virtual pion
field of the nucleon is Compton scattering with real and virtual
photons.  This is seen in the expression for the electric and
magnetic polarizabilities in chiral perturbation theory (ChPT)
which diverge as $1/m_{\pi}$\cite{CS-ChPT} indicating pion
dominance.  A new development in this field has been the
completion of a virtual Compton scattering (VCS) experiment at
$Q^{2} = 0.057~{\rm GeV}^{2}$ at Bates <Miskimen>\cite{VCS}. The
results are in agreement with heavy Baryon ChPT to order
$O(p^{3})$\cite{VCS-ChPT}. An interesting aspect of the $Q^{2}$
dependence of the electric polarizability $\alpha(Q^{2})$ is the
extraction of its RMS radius $r_{\alpha}^{2} = 2.16 \pm 0.31~{\rm
fm}^{2}$, which is in reasonable agreement with the ChPT
prediction of  1.7 fm$^{2}$ and  far larger than the RMS electric
radius of the proton $r_{p}^{2} =  0.757 \pm 0.014~{\rm fm}^{2}$.
This also indicates a pion-cloud dominance for the electric
polarizability of the proton.

Based on the  spontaneous hiding of chiral symmetry, which leads
to the strong p wave coupling of pions to nucleons, it  is
generally agreed that pionic effects play an important role in
nucleon structure.  This is reinforced by a large body of
empirical evidence.  As was discussed in the previous section,
the pionic contribution is only a qualitative concept and cannot
be made model independent. This was seen in particular in the
discussion of the pionic contribution to nucleon form factors.
There  the definition of pionic effects are  not the same as in
the case of dispersion calculations <Meissner>\cite{disp-pion}
and in the Friedrich-Walcher empirical definition \cite{FW}.

\subsection{Models for the $\gamma N \rightarrow \pi N$ and $e N
\rightarrow e' \pi N$ Reactions}

The ideal way to compare theory and experiment for  the
$\gamma^{*} N \rightarrow \pi N$ reaction would be to start with
QCD and compare the experimental observables with the
predictions. The first step towards this approach   has been
taken with a chiral effective field theory (ChEFT) calculation to
next to leading order (NLO)
<Pascalutsa-Vanderhaeghen>\cite{pasc}. This work also estimated
the theoretical errors due to the next order (NNLO). Within these
estimated errors the calculations are in reasonable agreement with
experiment. Unfortunately, these errors are relatively large so
that these next order calculations are required for a quantitative
comparison between theory and experiment. The other QCD based
calculations, lattice\cite{Dina} and  ChEFT
<Gail-Hemmert>\cite{gh} calculate the multipoles, so that these
need to be extracted from experiment.

The $\Delta$ resonance ($J = I =3/2$)  is at a center of mass
energy of $\simeq$ 1232 MeV, and has a width $\Gamma \simeq $100
MeV \cite{PDG}. It decays $\simeq$ 99\% to the $\pi N$ channel and
$\simeq$ 1\% to the $\gamma$ channel. Its shape and width are
dominated by the $\pi N$ interaction. Both photo- and electro-pion
reactions are strongly linked to the $\pi N$  channel by unitarity
(Fermi-Watson theorem). The large width of the $\Delta$ has
important dynamical implications which must be taken into account
in proper dynamic calculations. Unfortunately in lattice
calculations with heavy pions, quark models and pion extensions,
the $\Delta$ is treated as an excited state  and not part of the
$\pi N$
continuum\cite{Dina,Eisenberg,cloudy2,cloudy3,Sirca,Buchmann,Buchmann2}.
This has the important consequence that the  $\gamma^{*} N
\rightarrow \Delta$ transition is limited by angular momentum and
parity considerations to have three possible multipole
amplitudes. These are the dominant magnetic dipole M1 due to
quark spin flip and the electric and Coulomb quadrupole  E2, and
C2 amplitudes which signal the non-zero angular momentum
components in the nucleon and $\Delta$ states. This represents a
simplification which must be removed in a realistic analysis of
the data. What is measured is the $\gamma^{*} N \rightarrow \pi
N$  reaction initiated by real or virtual photons. Since this is
a continuum final state, there are  an infinite number of
possible multipoles. At low energies the practical number of
significant multipoles is limited to $\simeq$ kR where k is the
momentum of the outgoing pion and R is the nucleon size. For
energies of 1 GeV or less we expect contributions of less than 5
units of angular momentum. In addition we have to take into
account the fact that there are two final charge states possible
for either a proton or neutron target, which is conveniently
characterized in terms of isospin.

In electro-pion production notation (see \cite{drechsel_tiator}
for details) the three allowable $\Delta$ resonance amplitudes
M1, E2, and C2 are $M_{1+}^{3/2}, E_{1+}^{3/2}, S_{1+}^{3/2}$
where the letter ($M, E, S$) stands for magnetic, electric, and
scalar.. The subscript (1+) stands for the  angular momentum L=
1, the plus sign means  that the total angular momentum $J = L +
1/2$=3/2. The superscript is the isospin I
=3/2\cite{drechsel_tiator}. All other multipoles  are background
coming primarily from Born terms and from the tails of higher
resonances <Drechsel-Tiator>.  The multipoles are complex numbers
which are functions of $W$ and $Q^{2}$. The Fermi-Watson theorem
states that the phase  angle is the same as in $\pi N$ scattering
(a function of W only). This depends on unitarity, isospin
conservation, and the fact that only the $\pi N$ and $\gamma$
channels are open so that it is not strictly valid above the two
pion threshold. The small deviations of the Fermi-Watson theorem
due to isospin breaking (e.g. by the mass difference of the up
and down quarks) can be removed if the precision of the
experiments warrants\cite{AB-FW}.

Reaction models of  $\gamma^{*} N \rightarrow \pi N$  play a
central role in the determination of the multipole amplitudes.
None of the experiments performed to date has a sufficient number
of polarization observables to perform a model independent
determination of the multipoles and so models have to be utilized
to extract the multipoles of interest. Typically, experiments
extract the resonant multipoles $M_{1+}^{3/2}, E_{1+}^{3/2},
S_{1+}^{3/2}$ from experiment by fitting them to the data and
assuming that the background (Born) amplitudes are properly
described by the model. In some cases the extraction uses several
models to test this hypothesis <Stave~{\it et~al.}>. The
calculations that have been used in this way are the pion cloud
dynamic models   of Sato and Lee\cite{sato_lee} and
DMT\cite{dmt}. There is the phenomenological MAID
<Tiator-Kamalov>\cite{maid} and closely related calculation of
Aznauryan\cite{Azn03}, and finally the multipole fitting SAID
program <Arndt~{\it et~al.}>\cite{said}. For a fuller discussion
of these calculations see <Drechsel-Tiator>. At the workshop a
model independent method to extract the multipoles was
presented<Stiliaris-Papanicolas>.

\subsection{High Energy Probes and Generalized Parton Distributions}

Inclusive deep inelastic lepton scattering (DIS) was used to
determine that spin 1/2 confined quarks make up the proton and to measure their distribution functions (for a review of
the data see \cite{PDG}). From experimental integrals of the
distribution functions it was determined that the quarks carry
only $\simeq$ 50\% of the proton momentum. Later  DIS
measurements with polarized beams determined that only $\simeq$
25\% of the proton spin comes from the valence quarks
\cite{spin-review}. There are concerted efforts to measure the
spin contribution of the $\bar{q}q$ quark sea, the  gluons and
the angular momentum of the quarks with high energy leptons, and
polarized proton scattering at RHIC. These promising experimental
studies are just beginning. It is believed that a significant
amount of the proton spin is carried by the angular momentum of
the quarks and by the gluons\cite{spin-review}. These high energy
studies are complimentary to the main focus of this workshop,
which has  been on the long range part of the non-zero angular
momentum amplitudes in the proton and $\Delta$. As we have seen,
these are largely due the pion (more generally $\bar{q} q$)
contribution. We believe that the future work, using high energy
probes, will map out the short range, quark, angular momentum
amplitudes and connect the short and long range contributions. It
may be possible to identify  the quark structure of the pion
cloud and connect parton distribution information with form
factors through sum rules.

In the past few years there have been developments in mapping out
the structure of the proton through Generalized Parton
Distributions (GPD) which can be measured in deep  virtual
Compton scattering (DVCS) and hard exclusive meson production
<Vanderhaeghen>. The  GPDs contain information about both the
spatial and momentum distribution of the quarks.  A Fourier
transform of the GPD yields a tomographic view of the nucleon,
i.e. the simultaneous distribution of quarks with longitudinal
momentum x and transverse position b <Kroll>.

\begin{figure}
\includegraphics[height=0.2\textheight]{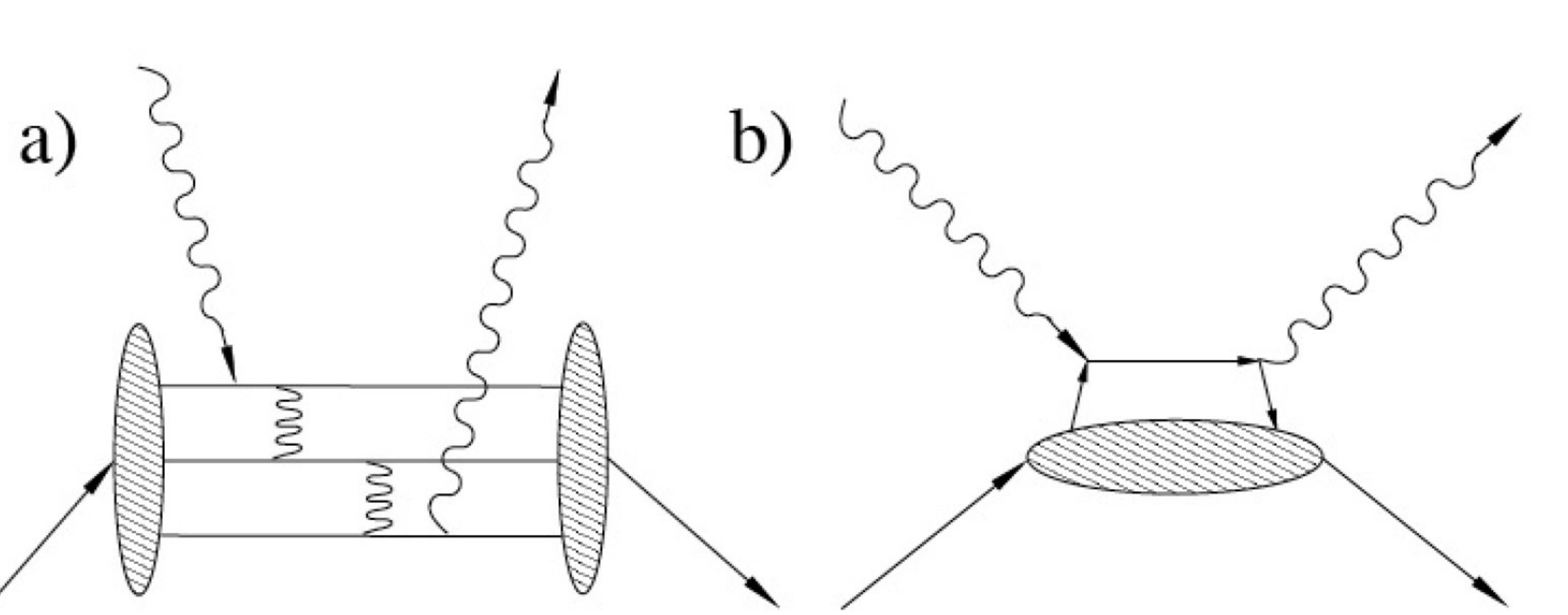}
\caption{Diagrams for the $N \to N$ and $N \to \Delta$  processes
for real and virtual Compton scattering at high energies. a)
Momentum shared by many quarks due to gluon exchange; b) the
handbag diagram- momentum absorbed by a singe quark.The diagram
with the incoming and outgoing photon lines crossed is not shown.
Figure from\cite{JLab-DVCS}.} \label{fig:dvcs}
\end{figure}

In Compton scattering at high energies, two reaction mechanisms,
shown in Fig.~\ref{fig:dvcs}  have been proposed. The left hand
diagram shows the initial hard real or virtual photon being
absorbed by many quarks with virtual gluons mediating the
interaction. The right hand "handbag" diagram shows the incident
photon being absorbed by one quark. There have been two recent
JLab experiments which show the dominance of the handbag
diagram. One involves the polarization transfer from circularly
polarized photons at $s =6.9$ GeV$^{2}$ and $t = -4.0$
GeV$^{2}$\cite{JLab-DCS}. The second JLab experiment confirmed
that the handbag diagram can be extracted from experiment in a
deep virtual Compton scattering  DVCS experiment at modest
$Q^{2}$(1.5 to 2.3 GeV$^{2}$)\cite{JLab-DVCS}. In the "handbag"
diagram the struck quark propagates after (or before) emitting
the outgoing photon so that in contrast to elastic lepton
scattering, which is primarily mediated by a single photon
exchange, DVCS can provide information about the angular momentum
of the struck quark. The "handbag" amplitude factorizes into a
hard and soft part, shown as the lower blob which represents the
generalized parton distributions.  We anticipate a large amount
of work in this field in the coming decade.

The GPDs have been parameterized by using models and the
appropriate integrals over the GPDs (one of which gives the
nucleon form factors) <Vanderhaeghen, Kroll>\cite{GPD-FF,
Ji-DVCS}. Figure~\ref{fig:dvcs} shows the DVCS for either the
nucleon or the $\Delta$ as the final state. By taking the proper
integrals of the GPDs for the N $\rightarrow \Delta$ transition,
and by utilizing large $N_{c}$ calculations, the form factors of
the $\gamma^{*} N \rightarrow \Delta$ transition can be obtained
and are in reasonable  agreement with experiment (see Fig.~12 of
<Vanderhaeghen>). Another way to look at the information obtained
with GPDs is through the Wigner phase space distribution which
links the spatial and momentum quark correlations\cite{Ji-Wigner}.

Observation of non zero quark angular momentum in the proton has
been demonstrated  by the HERMES collaboration <Rith>. They have
observed single spin asymmetries  in semi-inclusive deep
inelastic scattering DIS  for transversely polarized proton
targets. They measured non-zero Sivers  distribution functions
for semi-inclusive DIS with positive pions and kaons. The Sivers
function describes the correlation between intrinsic quark
transverse momentum $p_{T}$ and transverse nucleon spin and
requires  orbital angular momentum in the nucleon to be non-zero
<Rith>.

The field of high energy deep inelastic scattering with
polarization degrees of freedom has great potential to measure L >
0 angular momentum amplitudes in the nucleon. We are presently
at the beginning of this very interesting development.

\section{Experimental Measurements and Interpretation}
\label{expLandscape} The experimental landscape concerning the
investigation of shape of hadrons, until quite recently, has been
to a large degree dominated by the quest for resonant quadrupole
amplitudes in the \GNdelta transition. It is evident from the
proceedings of this workshop that the situation is rapidly
changing: other reactions have been suggested and some have been
explored theoretically. Issues concerning the study of form
factors and of GDPs have bearing on the issue and the theoretical
framework is currently being developed. The possibility of
studying the \GNdelta transition in  neutrons or in a nuclear
target has been raised and it may become technically feasible in
the near future.  In addition to the formidable technical
difficulties that such a program faces, the theoretical framework
to extract the above mentioned and admittedly important physics,
needs to be further developed.

\subsection{ The \GNdelta ~Channel}
As commented earlier the \GNdelta~ transition currently offers the
only quantitative experimental  test of theories and models that
incorporate mechanisms which predict  hadron deformation in
general and that of the proton in particular. This avenue
developed in the late eighties~\cite{Excited Baryons1988}  has
been pursued and refined both theoretically and experimentally
over the last thirty years. The experimental landscape as it
emerged in the intervening years can be classified according the
reaction channel probed and on how exclusive it is. The
$\Delta^{+}(1232)$ can be excited by real or virtual photons and
it can decay~\cite{PDG2006}  by $\pi^{\circ}$ (66\%) or $\pi^{+}$
(33\%) or by gamma decay ($0.56\pm0.04$\%):
$$\gamma^{*} p \rightarrow \Delta^{+}(1232)\rightarrow p \pi^{\circ}~~
(66\%)$$
$$\gamma^{*} p \rightarrow \Delta^{+}(1232)\rightarrow n \pi^{+}~~
(33\%)$$
$$\gamma^{*} p \rightarrow \Delta^{+}(1232)\rightarrow p \gamma~~~~
(0.56\%)$$ \noindent where $\gamma^{*}$ denotes the real (often
polarized) or virtual photon inducing the excitation.  The first
two channels have been extensively explored while the third,
involving the small gamma decay branch, has been studied with
real compton scattering measurements (RCS). Virtual compton
scattering measurements (VCS)on the $\Delta^{+}(1232)$ are just
beginning to emerge with the dual aim of either mapping the
polarizabilities at high missing mass~\cite{DeltaVCS_a} and/or
the issue of deformation~\cite{sparveris_vcs}.

It is possible to classify nucleon resonance photoproduction
experiments by the exclusivity of the reaction channel studied.
We classify the \GNdelta experiments (and nucleon resonance
studies in general) as "first", "second" and "third" generation
using such a criterion. We term "first generation" experiments
those that established the reaction mechanisms and obtained the
first data. We call second "generation experiments" those that
have used high quality beams and polarization degrees of freedom
either in the entrance or the exit channel. Finally we call
"third generation" experiments those that employ double
polarization degrees of freedom: either both in the entrance and
exit channels or equivalently polarized beams and targets. The
first generation experiments were conducted in the late sixties
and early seventies before the issue of deformation was even
raised- for an excellent review see the monograph
by Amaldi, Fubini and Furlan~\cite{AFF 1979}.  These measurements were performed
with low duty factor accelerators (DESY, NINA,CEA), low quality
beams and with experimental equipment not designed to address such
refined questions. The data that emerged, as far as the issue of
deformation was concerned, were inconclusive, and limited by big
systematic uncertainties and poor statistics, but they did provide
valuable guidance on the design of the second generations
experiments~\cite{CNP1989}. The second generation experiments were
obtained by a newer generation machines (Brookhaven, Bates, MAMI,
CEBAF) with optimized or especially designed equipment, such as
the OOPS spectrometer at Bates, and in general with polarized
beams. Third generation experiments are now beginning to emerge;
they have been conducted primarily with real photons (polarized
and tagged, impinging on polarized targets). Electroproduction
experiments with polarized targets are particularly difficult
with only one measurement reported in the literature using the
internal target facility at NIKHEF ~\cite{bartsch}, of low
statistical accuracy. This experiment should be viewed as a
valuable demonstration of an important technique. Recently the BLAST collaboration at Bates has measured the inclusive double-polarized response from hydrogen spanning the $\Delta$ region from 1100-1400 MeV and $Q^{2}$ from ~0.08 to 0.3 GeV$^{2}$  with small systematic errors for the double asymmetry. This will allow them to obtain a good  extractions of the CMR and EMR\cite{Kohl}. The recently
reported JLab Hall A experiment~\cite{kelly} which presented high
quality extensive recoil polarization measurements using
polarized beams is truly third generation experiment which both
demonstrated the feasibility of the technique, the precision that
can be achieved and the rich physics output that can emerge.
Unfortunately, no similar  follow up measurements are scheduled
for the immediate future.

In general, in the real photon sector the "second generation"
experiments are completed and analyzed and the era of "third
generation" experiments is about to begin in earnest, in view of
the important instrumentation initiatives~\cite{kotula} at Mainz
and at Bonn. In the electroproduction sector, data are still
emerging from second generation experiments while experiments of
third generation are just beginning to emerge. Jlab and MAMI C
have optimal beams and detection systems for the pursuit of this
program which is far from being exhausted.

The coincident $p(\vec{e},e^{\prime}\pi)$ cross section in the
one-photon-exchange-approximation can be written
as~\cite{drechsel_tiator}:

\begin{eqnarray}
\frac{d\sigma}{d\omega d\Omega_{\rm e} d\Omega^{\rm cm}_{\pi}}
    =
    \Gamma_{\rm v}~\sigma_{\rm h}(\theta,\phi)   \\ \nonumber
    \sigma_{\rm h}(\theta,\phi)=
    \sigma_{\rm T} + \varepsilon \sigma_{\rm L} +
    \sqrt{2\varepsilon(1+\varepsilon)} \sigma_{\rm TL} \cos\phi\\ \nonumber
    + \varepsilon \sigma_{\rm TT}\cos 2\phi
+ h p_{\rm e}\sqrt{2\varepsilon(1-\varepsilon)}\sigma_{\rm
TL^{\prime}}\quad , \label{eq:xsection}
\end{eqnarray}

\noindent where $\Gamma_{\rm v}$ is the virtual photon flux, $h =
\pm 1$ is the electron helicity, $p_{\rm e}$ is the magnitude of
the longitudinal electron polarization, $\varepsilon$ is the
virtual photon  polarization, $\theta $ and $\phi$  are the pion
CM polar and azimuthal angles relative to the momentum transfer
$\vec{q}$, and $\sigma_{\rm L}$, $\sigma_{\rm T}$, $\sigma_{\rm
TL}$, and $\sigma_{\rm TT}$ are the longitudinal, transverse,
transverse-longitudinal, and transverse-transverse interference
cross sections, respectively~\cite{drechsel_tiator}.  In the case
of $p(\vec{\gamma},\pi)$ the longitudinal responses are absent.
The small quadrupole resonant quadrupole multipoles
~($E^{3/2}_{1+}$)  and ~($S^{3/2}_{1+}$) have been  investigated
using the interference responses $\sigma_{LT}$ and $\sigma_{TT}$,
where the small amplitudes are exposed by interfering with the
dominant M1 amplitude.

\subsubsection{Real Photon Measurements}
Precision measurements with polarized tagged photons performed at
Mainz and Brookhaven (LEGS) in the late nineties have converged
at the level of asymmetries resulting in a resonant EMR$ = {\rm
Im} E_{1+}^{3/2} /  {\rm Im} M_{1+}^{3/2}$ of $(-3.0 \pm 0.3)\%$
\cite{blanpied} and $(-2.5 \pm 0.3) \%$ \cite{beck}.  A number of
theoretical calculations are in good agreement with the
experimentally derived EMR. However, important discrepancies still
persist in the measurements of the two labs in the detailed
angular and energy distributions. Both the $(\gamma, \pi^{\circ})$
and the $(\gamma, \pi^{+})$ channels have been studied
extensively. $(\gamma, \gamma)$ (RCS) has also studied~\cite{RCS},
where the resonance pion-photoproduction amplitudes were evaluated
leading to the multipole E2/M1 ratio EMR(340 MeV) =$(-1.6 \pm
0.4_{(stat+syst)} \pm 0.2_{(model)}$)\%, in reasonable agreement
with the photopion measurements.

\begin{figure}[h]
\includegraphics[width=1.00\textwidth,height=0.60\textwidth,angle=-00]{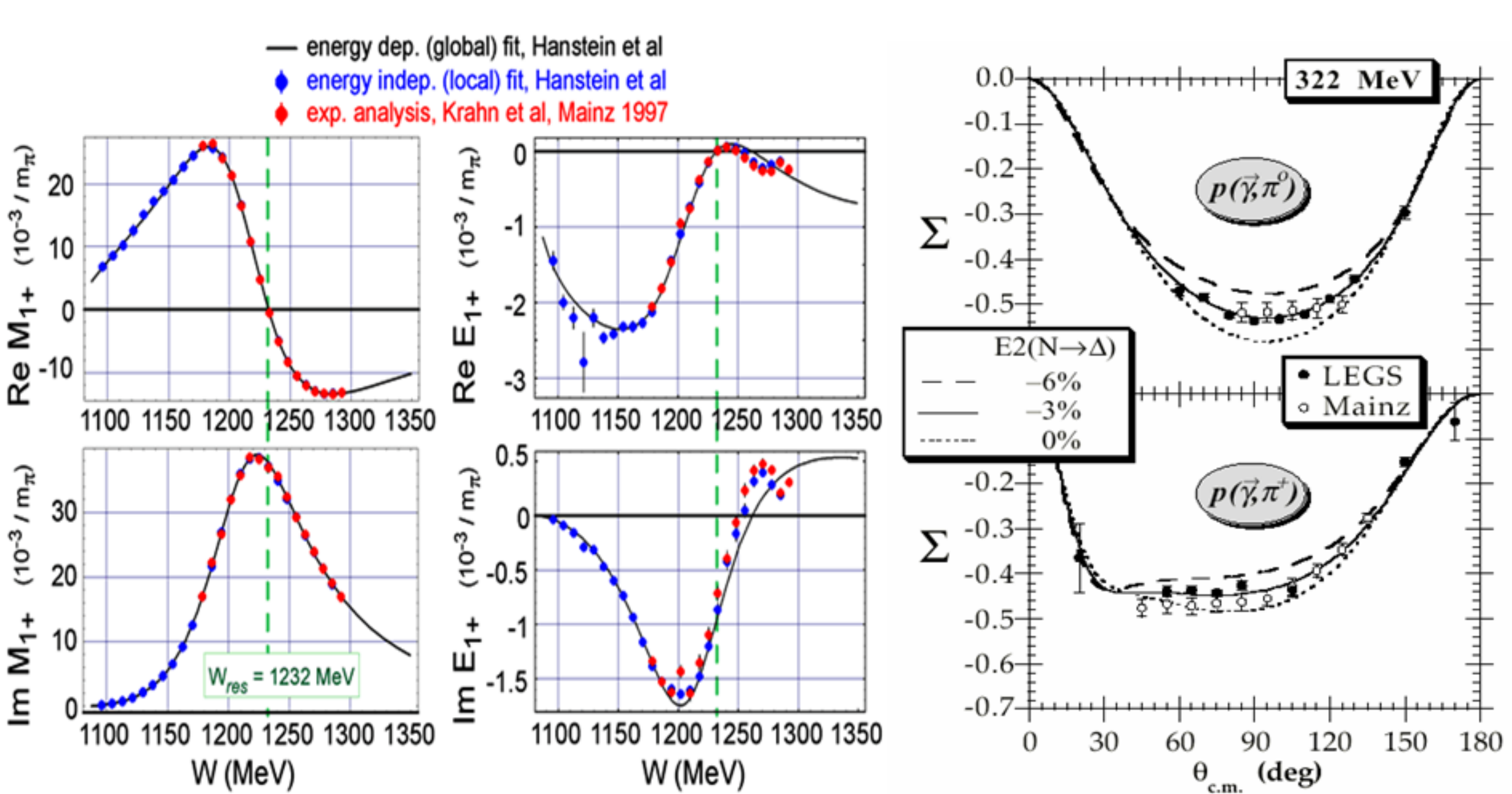}
\caption{Measurements from MAMI and LEGS have yielded precise
measurements of the resonant quadrupole amplitude at the photon
point. A most sensitive probe is the polarization asymmetry
$\Sigma=(\sigma_{\|}-\sigma_{\bot} )/(\sigma_{\|}+\sigma_{\bot} )$
which has been measured precisely at both MAMI and LEGS.The
derived multipoles from the Mainz cross sections, yield an
accurate measurement of EMR.}

\label{fig_realphtons}
\end{figure}

The "deformation" signal in the real photon sector comes from the
study of the  $E_{1+}^{3/2}$ (transverse quadrupole) multipole.
The desired signal is particularly hard to isolate because the
transverse channel is overwhelmed by the $M_{1+}$  amplitudes and
contaminated with other background processes of similar magnitude.
In this sense, the $E_{1+}^{3/2}$ appears in next to leading order
(NLO) in photoproduction. The impressive results from LEGS and
MAMI are a tour de force of experimental finesse: they make the
small quadrupole amplitude visible,as it can be seen in the right
panel Figure~\ref{fig_realphtons}, and they extracted with small
systematic uncertainty  by measuring the polarization asymmetry
$\Sigma=(\sigma_{\|}-\sigma_{\bot} )/(\sigma_{\|}+\sigma_{\bot}
)$ which is measured by flipping the polarization of the tagged
photon beam parallel $(\|)$ and perpendicular  $(\bot)$ to the
beam direction. It is exactly for this reason that we have
convergence at the level of asymmetry but not at the level of
cross sections. Analysis of the MAMI $(\gamma, \pi^{+})$ and
$(\gamma, \pi^{\circ})$ data, yields the impressive results shown
in the left panel of Figure~\ref{fig_realphtons}. It is obvious
from the figure that the derived results heavily depend on the W
dependence of the cross section. It is also quite revealing that
the $E_{1+}^{3/2}$ mutlipoles have a striking non resonant shape,
a manifestation of the complicated processes that contribute to
this channel, as discussed in <Drechsel>.

The situation concerning the \GNdelta transition in the photon
sector has remained stable, without experimental results reported
to change this picture in the last five years. More recent
analysis~\cite{BRAG01} and data<Kotula> which give EMR =$-2.74
\pm 0.03_{(stat)} \pm 0.3_{(syst)}$, confirm the EMR values
of~\cite{blanpied,beck}. However, in the closely related areas of
threshold pion production~\cite{Merkel06} and in the measurement
of the magnetic dipole~\cite{kotula, kotula02} of the
$\Delta^{+}(1232)$ the very precise results that emerged are
testing and providing valuable guidance to theory and
phenomenology that is common to both. The recent installation of
the crystal ball at MAMI and of the crystal Barrel in Bonn, have
brought into existence new very powerful tools and as a result
new more precise data and results can now be expected.

\subsubsection{Electroproduction measurements}
Electron scattering experiments offer a far richer field of
study, in comparison to real photons, since in addition to the
transverse responses the longitudinal responses are accessible
which are sensitive to leading order to the $L_{1+}^{3/2}$. In
addition to the W dependence the $Q^{2}$ evolution of the various
responses can be studied as well. The $Q^{2}$ dependence offers,
as is well known, the ability to distinguish between large scale
and short scale process, of particular value in the quest to
distinguish between the effects of the mesonic cloud from those of
the quark core. These advantages are however hard to realize and
time consuming due primarily to the numerous measurements needed
to cover the widest possible range of momentum transfers. This
explains the time lag with which electron scattering experiments
have appeared: while the real photon measurements appear to have
converged already at approximately around the year 2000, the
electroproduction experiments are only now reaching a similar
stage of maturity.

\begin{figure}[h]
\includegraphics[width=0.8 \textwidth,height=0.65\textwidth,angle=-00]{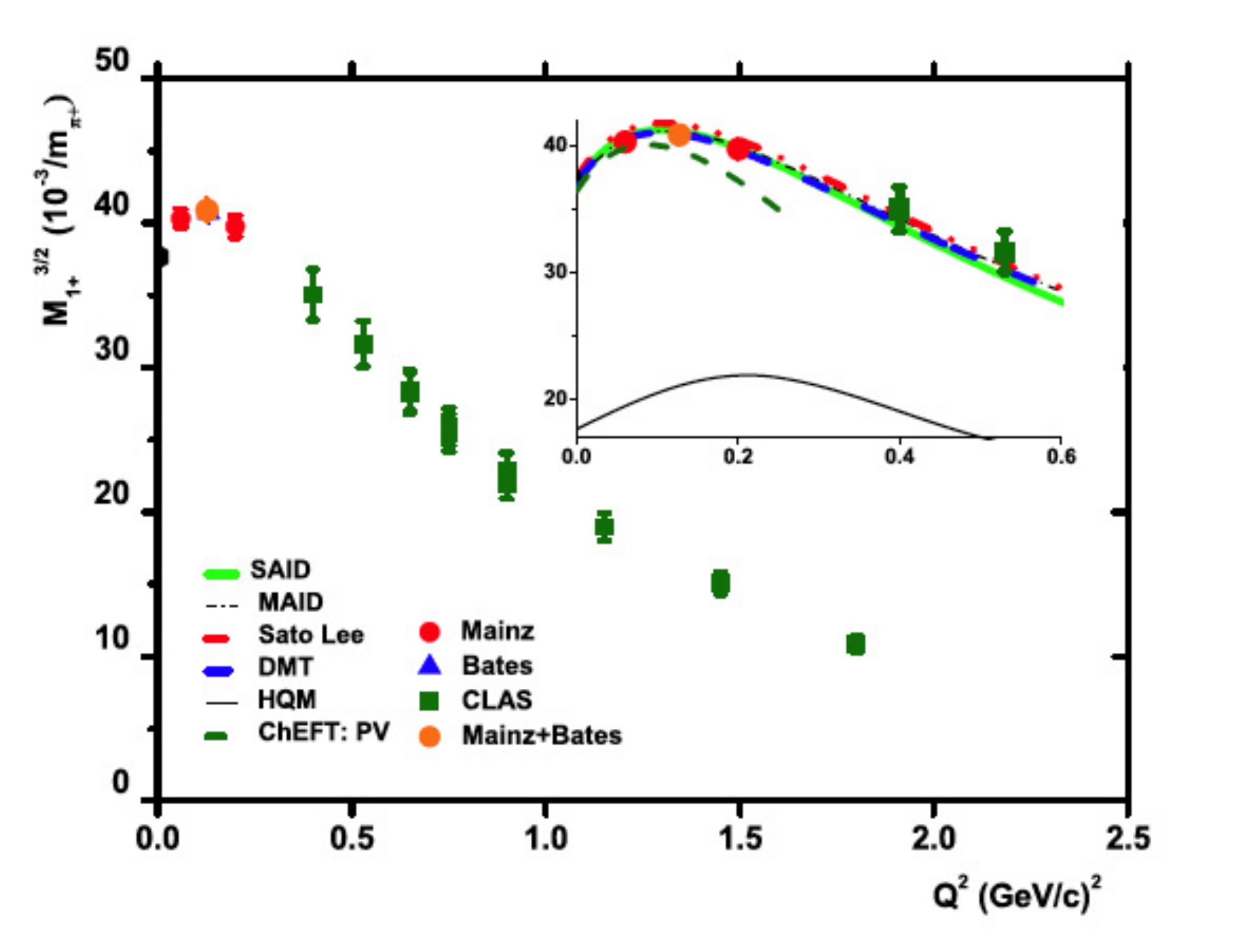}
\caption{Experimentally derived M1 values compared to lattice and
EFT results. The derived multipole ratios are generally shown
without the model error that is of the order or larger than the
depicted experimental error.} \label{fig:fig_M1}
\end{figure}

Results from several groups which are consistent and converging
have been reported in ~\cite{nstar2001} and in this volume. These
experiments \cite{warren, mertz, pospischil, kalleicher, frolov,
joo, stave} at Bates, Bonn, MAMI an JLAb have mapped the momentum
transfer range from $Q^{2}$= $0.06$ to $6.0$ \gevc with high
precision in a limited number of sensitive observables. Earlier
discrepancies between measurements from different labs have been
resolved: There are no known outstanding  discrepancies of
relevance in electroproduction data, an achievement of the last
few years. There are discrepancies at the extracted EMR and CMR
values (which are not experimental observables)which are primarily
due to the methodology that is used in extracting multipoles, as
it will be discussed in the next section.

\begin{figure}[h]
\includegraphics[width=0.9\textwidth,height=0.60\textwidth,angle=-00]{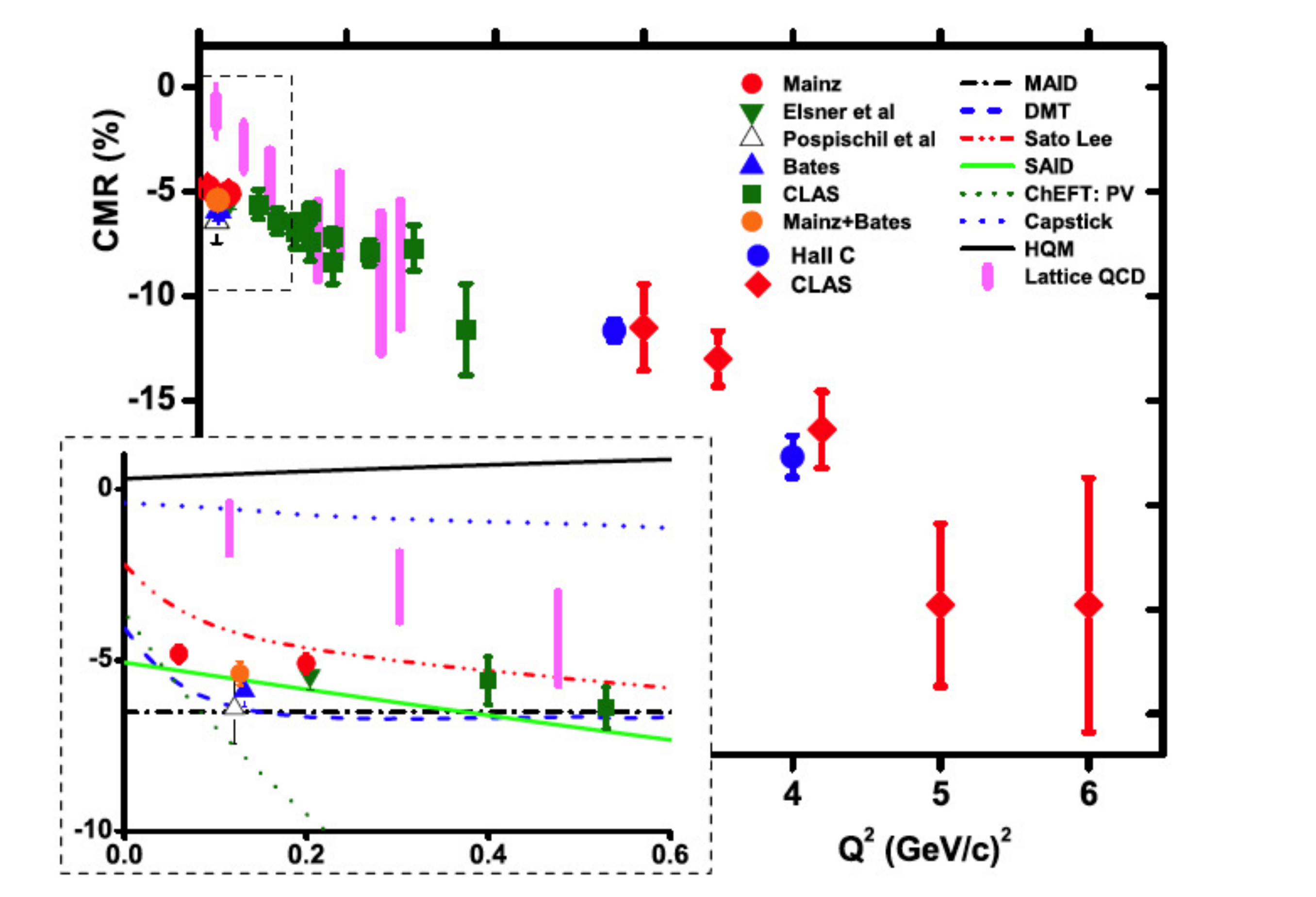}
\caption{ Experimentally derived CMR values compared to lattice
and EFT results. The derived multipole ratios are generally shown
without the model error that is of the order or larger than the
depicted experimental error.} \label{fig:CMR}
\end{figure}

Starting from the experimental observables two methods have been
used and are reported in the literature for  extracting multipole
amplitudes: a) In the Truncated Multipole Expansion (TME)
approximation most or all of the non resonant multipoles are
neglected (e.g. see~\cite{kalleicher, frolov}) assuming that at
resonance only the resonant terms contribute significantly, and
b)In the Model Dependent Extraction (MDE) method  a
phenomenological reaction framework with adjustable quadrupole
amplitudes is used to perform a model extraction (e.g.
see~\cite{frolov, mertz, stave}). It is assumed that the reaction
is controlled at the level of precision required for the
disentanglement of the background from the resonance. Clearly the
MDE method is superior, given the sophistication that
phenomenological models have achieved in describing the  data,
and as a result the TME method needs only be used  to set a base
line for the importance of the resonant terms and for the
pedagogical value it offers.

\begin{figure}[h]
\includegraphics[width=1.00\textwidth,height=0.70\textwidth,angle=-00]{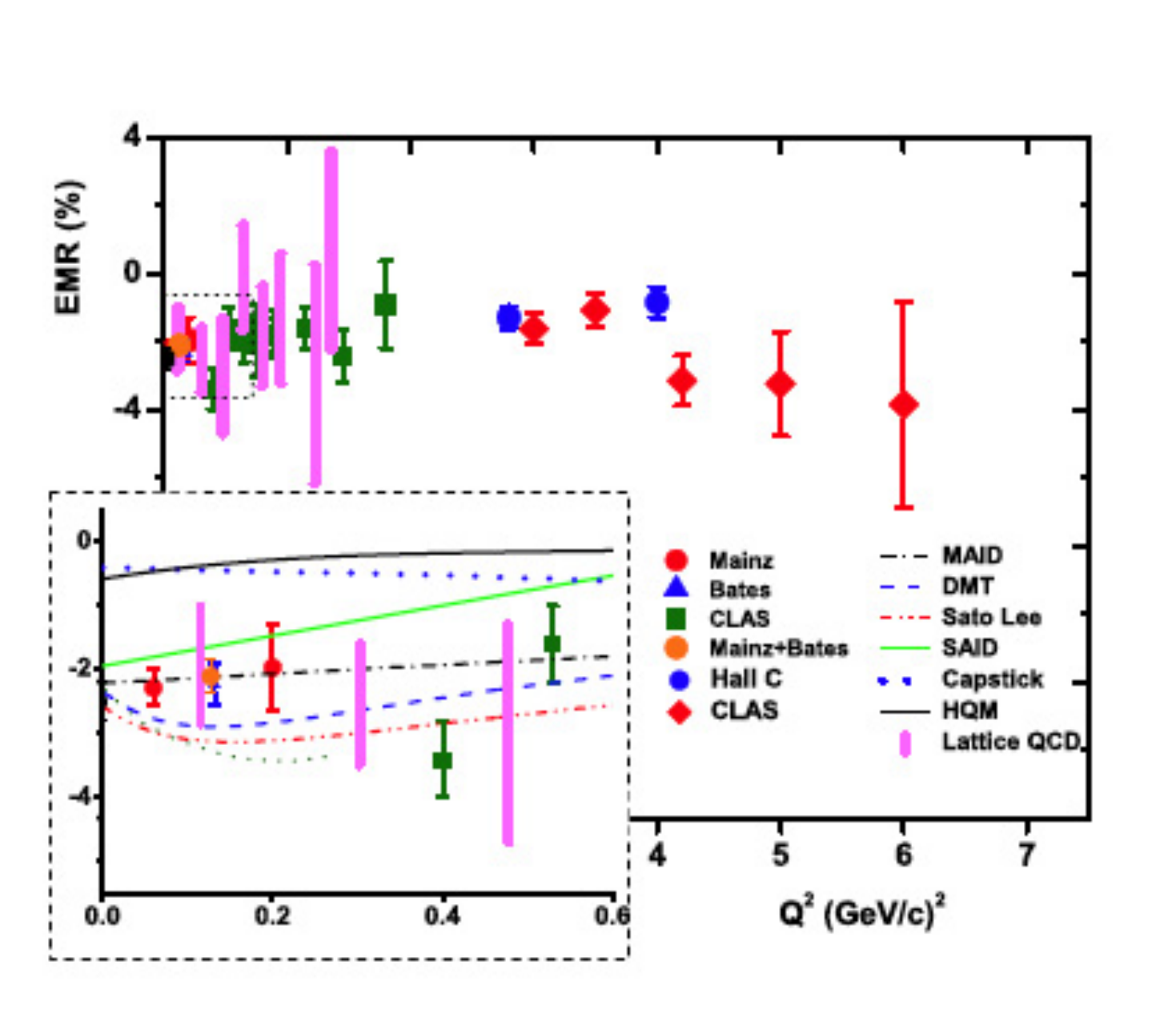}
\caption{  Experimentally derived EMR values compared to lattice
and EFT results. The derived multipole ratios are generally shown
without the model error that is of the order or larger than the
depicted experimental error.} \label{fig:EMR}
\end{figure}

In the recent electroproduction experiments which almost
invariably are carried out with polarized beams, the TL and the
TL$^{\prime}$ (transverse-longitudinal) response functions are
the real and imaginary parts of the same combination of multipole
amplitudes. In the S\&P ($L_{max}=1$) wave approximation they can
be written as~\cite{drechsel_tiator}:

\begin{eqnarray}
    \sigma_{\rm TL}(\theta) &\simeq&
        - \sin\theta {\rm Re}[ A_{\rm TL} +B_{\rm TL}
\cos\theta]
    \\
    \sigma_{\rm TL'}(\theta) &\simeq&
        \sin\theta {\rm Im}[ A_{\rm TL} +B_{\rm TL}
\cos\theta]
    \nonumber \\
    A_{\rm TL} &\simeq& -L_{0+}^{*}M_{1+} \nonumber
\\
    B_{\rm TL} &\simeq& -6L_{1+}^{*}M_{1+}\nonumber
\label{eq:TL}
\end{eqnarray}

\noindent

The two responses are particularly valuable because $\sigma_{\rm
TL}$ is most sensitive to the presence of resonant quadrupole
amplitudes while $\sigma_{\rm TL'}$ is particularly sensitive to
the background contributions, thus providing information on the
two key experimental issues being explored and which need to be
controlled independently. The importance of background is clearly
seen in the W behavior of the responses ~\cite{mertz} and the
non- vanishing recoil polarization $P_n$ \cite{warren,
pospischil}, which bear close resemblance to the fifth response.

The $TT$ (Transverse- Transverse) response which is sensitive to
the electric quadrupole amplitude, and which is the primary source
of information for the EMR at the photon point was only recently
isolated for the first time at non-zero $Q^2$ has in the last five
years being vigorously being pursued at Bates, Jlab and
MAMI~\cite{sparveris_SOH, cole_SOH}.   The response functions
\sT\ and \sTT\ contain the term $\Re[E_{1+}^*M_{1+}]$ but also
the dominant term $|M_{1+}|^2$. The influence of the dominant
$|M_{1+}|^2$ term can be diminished by measuring the following
combination of the \sT\ and \sTT\ responses:
\begin{eqnarray}
\sigma_{E2}(\thtpq) = \sT(\thtpq) + \sTT(\thtpq)-\sT(0) &\simeq&
                                                      \nonumber \\
2\Re[ \Ezpc(3\Eop + \Mop - \Mom)](1-\cos\thtpq)
                                                      \nonumber \\
- 12 \Re[E_{1+}^*(M_{1+} - \Mom)] \sin2\thtpq \label{eqn:vary1}
\end{eqnarray}
The term of interest $\Re[E_{1+}^*M_{1+}]$ is  enhanced by a
factor of twelve (12) while the leading term ($|M_{1+}|^2$ ) is
eliminated~\cite{cnp}! This has been used as as a tool to extract
a most precise measurements of EMR\cite{sparveris,sparveris2}.

As the $\gamma^{*} N \rightarrow \Delta$ data became more
accurate it was generally recognized that the quark model
predictions did not agree with the data. In particular the
dominant M1 matrix element was predicted to be $\simeq$ 30\% low
and the E2 and C2 amplitudes are generally at least an order of
magnitude too small and often of the wrong sign as shown in
Figs.~\ref{fig:CMR} and \ref{fig:EMR}. As was discussed in the
introduction it was soon realized that the pion cloud needed to
be added to quark models. This failure  is to be expected since
the quark model does not respect chiral symmetry whose dynamic
breaking leads to a strong, non-spherical, pion cloud around all
hadrons\cite{amb}. It was shown by the calculations of the
Sato-Lee~\cite{sato_lee} and DMT~\cite{dmt}  models that most of
the strength of the responses (and the EMR and CMR) at very low
$Q^2$ values (below $\simeq 0.25$ $GeV2/c2$) arises on account of
the mesonic degrees of freedom. Finally this has been
theoretically confirmed by the chiral effective field theory
calculations <Gail-Hemmert, Pascalutsa-Vanderhaeghen>\cite{pasc,
gh}. The recent results from MAMI along with the earlier ones
from Bates ~\cite{sparveris_SOH}and the recent low $Q^2$
measurements form CLAS~\cite{cole_SOH}, give strong support to
this interpretation.

Figures~\ref{fig:CMR} and Fig.~\ref{fig:EMR} offer a recent
compilation of the current status of CMR and EMR as a function of
$Q^2$. It can be observed that both EMR and CMR  are small and
negative in the region where they have been measured. At
asymptotic values of $Q^2$ helicity conservation~\cite{CP88}
requires that EMR $\rightarrow 1.0$ and that CMR $\rightarrow
$constant. Clearly this regime has not been reached. The upgrade
of CEBAF to 12 GeV  offers hope that the measurements will be
extended to higher $Q^2$, although this will pose significant
challenges in isolating the relevant partial cross sections and
even bigger ones in extracting the relevant amplitudes.

Finally, important is  the  investigation of the $H (\vec{e},e'p)
~\gamma$ channel which has not been exploited experimentally yet.
The  dispersion theory~\cite{Pa01} of the Mainz group, taken in
conjunction with the previous work of Vanderhaegen et
al.~\cite{va97} allows to addresses the physics of "deformation"
and of nucleon polarizabilities in the region above pion threshold
simultaneously. Very recent results from Mainz report the
extraction of polarizabilities~\cite{DeltaVCS_a} and the first
observation of VCS data sensitive to the resonant quadrupole
amplitudes ~\cite{sparveris_vcs}. Extraction of the quadrupole
amplitudes though VCS opens a new vista which can provide
important cross checks to the the derived results from the pionic
channel.

\subsection*{Sensitivity, Precision and Estimation of Uncertainties}

The progress achieved in the last few years in the experimental
investigation of the issue of deformation is truly outstanding.
In the \GNdelta investigations  data up to  $Q^{2}= 6.0$ \gevc are
in general characterized by small systematic error and high
statistical precision. The sensitivity of the data to the issue of
deformation had at about the same time been demonstrated, and as a
result, the research thrust shifted from the investigation of
whether the conjecture for deformation was valid to the
exploration of the mechanisms that cause it. The implications of
this shift had profound effect on the accuracy demanded from both
the experiment and the theoretical calculations. Investigating
the physical origin of deformation requires the comparison of
theoretical results with experimentally derived quantities, such
as multipole amplitudes at a much higher level of precision. A
number of models and recently lattice results are in qualitative
agreement with the data and amongst themselves. Understanding the
mechanisms that cause deformation requires  distinguishing among
successful descriptions that emphasize different aspect of the
same basic process. This in turn necessitates a precise statement
on the uncertainties that characterize both the experimental
results and the theoretical calculations to which they are
compared.

The situation described above is reminiscent of the "crisis" in
the analysis of elastic electron scattering data in the early
seventies, where the very precise electron scattering data that
were emerging could not be meaningfully compared with the
theoretical calculations of the time (primarily nuclear mean field
theories) at the level of charge densities. This was primarily
due to the lack of appropriate methodology which could enable the
quantification of uncertainties in the extracted densities, which
are not experimental observables. The resolution of that "crisis"
through the introduction of a "Model Independent" extraction of
charge densities led to a revolution in the field and to the
outstanding achievements in elastic and inelastic scattering of
the seventies and eighties.

In the quest for understanding of deformation through the \GNdelta
transition, a comparison is made to multipole amplitudes or their
ratios (EMR and CMR) which are also derived quantities not
experimental quantities. A quantification of the uncertainties
that characterize them is  difficult but essential.  Early
estimates showed that indeed the error uncertainty was in a
number of cases dominant. It is also the case that theoretical
calculations that predict them need to provide not only central
values  but an uncertainty as well. This is rarely done; even in
the case of lattice calculations, the quoted errors are
statistical not systematic and certainly not model. In the few
cases where this was done the associated uncertainty is quite
large.

The leading method of extraction of multipole amplitudes, the
Model Dependent Extraction (MDE), results in extracted values
that are biased by the model. The principal drawback of this
method is that the derived results are characterized by a had to
estimate model error, especially if a single model is
employed~\cite{joo,frolov,pospischil,bartsch,elsner,Ungaro} in
the extraction. An ansatz for estimating the model uncertainties
in the extracted multipoles has been proposed~\cite{cnp} and has
been used in a few cases ~\cite{sparveris,stave}. In this method
the same data are analyzed employing different models which
describe the data adequately, and attributing the resulting
spread in the extracted quantities to model uncertainty. A
critical assessment of the this approach in the case of the
$N\rightarrow \Delta$ transition and the resulting uncertainties
is presented in reference~\cite{SBN_SOH}. It is convincingly
demonstrated that the models currently available of considerable
sophistication provide though a MDE a convergent method of
extracting resonant amplitudes, at least at low $Q^{2}$.  The
role of the small non resonant amplitudes, which in MDE are fixed
by the model, is explored in detail. It is found that while each
one of them may be of little significance they collectively could
induce large correlations and that intrinsic model error is
similar in size to the model-to-model error. This thorough
investigation concludes that MDE, which constitutes the most
advanced available methodology, "without improvement in the
models, this is as far as the current data can take us".

\begin{figure}[h]
\includegraphics[width=1.00\textwidth,height=0.50\textwidth,angle=-00]{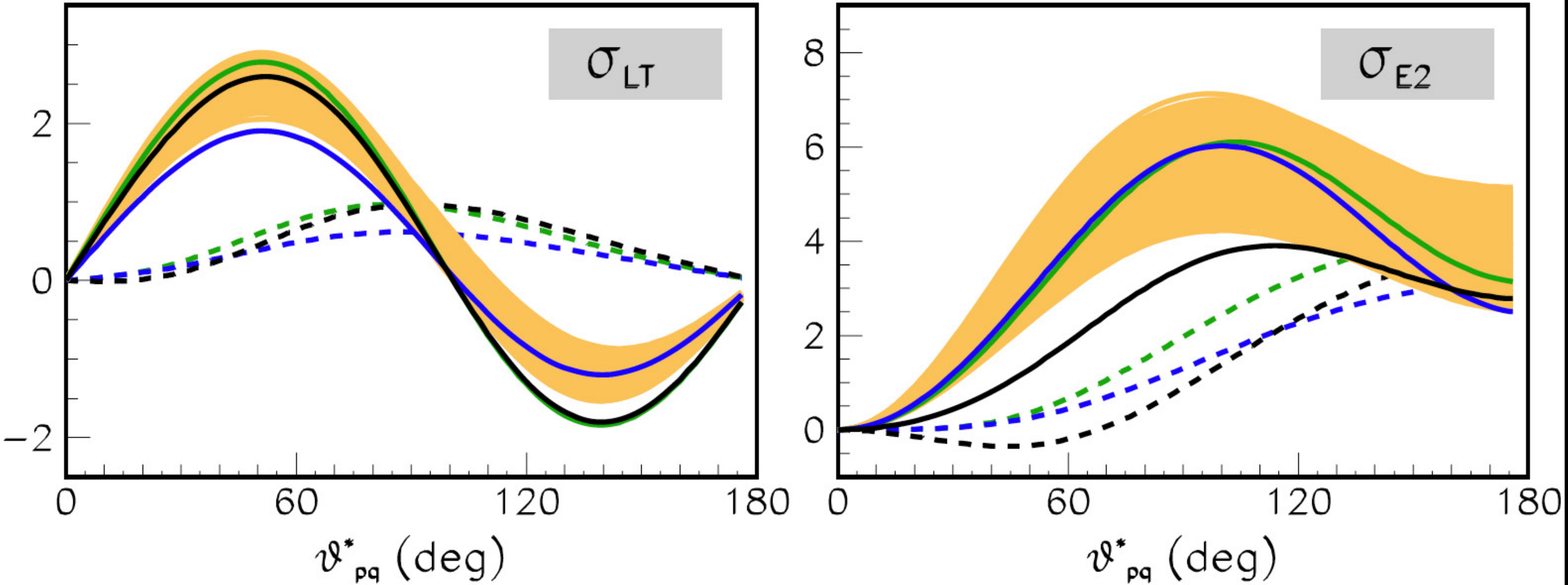}
\caption{The precise range of uncertainty that is allowed by the
$Q^{2}=0.127$ \gevc~ Bates and Mainz data for the \sLT~ and
\sEtwo~ responses is shown as a function of $\theta^*_{pq}$.
Current theoretical models (MAID in black, SL in blue and DMT in
green) predict this satisfactorily within a $2\sigma$ confidence
level. No "spherical" description (same color coding for the
various models  but with dashed curves) of the nucleon can
describe the data; they are excluded with high confidence.}
\label{fig errorbands}
\end{figure}

In an attempt to address the same fundamental issue a novel
method for extracting multipole information from experimental
nucleon resonance data was presented in the workshop
~\cite{AMIAS_SOH}. The Athens Model Independent Analysis Scheme
(AMIAS) is designed to be a rigorous, precise, and model
independent scheme and holds the promise to be a new tool for
nucleon resonance analysis with important consequences. The
analysis of the particular data discussed in the paper, is
demonstrated to yield new information on background amplitudes,
which MDE is incapable of accessing~\cite{SBN_SOH}. A very
distinct and promising aspect of the AMIAS method is its ability
to quantify the uncertainty of the extracted multipoles.

Results from AMIAS are shown  Figure~\ref{fig errorbands}
concerning the CMR sensitive \sLT~ and the EMR sensitive \sEtwo~
partial cross sections with the precisely defined $1\sigma$
(68\%) uncertainty. The experimentally allowed \sLT~ and \sEtwo~
partial cross sections, as constrained by the $Q^{2}=0.127$
\gevc~ Bates and Mainz data, are shown as function of
$\theta^*_{pq}$. They are compared with theoretical model
predictions that account for them. It is evident that the model
predictions (dotted curves) with resonant quadrupole amplitudes
set to zero, which amounts to spherical solutions, are excluded
with high confidence.  The "deformed" model predictions, if we
assume that they are characterized by negligible model error, are
in agreement at the $2\sigma$ level, which will allow their
characterization as reasonable. Differences among the various
model are visible, however commenting on their differences would
be far more meaningful, if their model error was known.  The
above comments not withstanding, the potential of AMIAS is
evident. The comparison that it offers demonstrates once again
that the assumption of sphericity for both the nucleon and the
$\Delta^{+}(1232)$ is incompatible with the data.

\section{Summary and Conclusion}
In this workshop the physical basis of the deviation of hadron
shapes, and in particular those of the nucleon and $\Delta$, from
spherical symmetry (non-spherical amplitudes) has been addressed
including the experimental methods and  results. At the present
time the only quantitative method experimentally accessible is
through the investigation of the $\gamma^{*} p \rightarrow
\Delta$ reaction, and the workshop focused on this. In addition
attention was payed to emerging new approaches using high energy
probes. A  promising method for which the theoretical framework
has been rapidly developing in recent years uses deeply virtual
Compton scattering and hard exclusive meson production to obtain
generalized parton distributions<Vanderhaeghen, Kroll>. In
addition semi-inclusive deep inelastic scattering scattering
experiments with transverse polarized beam and target also are
beginning to show the effect of non-zero quark angular momentum
(non-zero Sivers effect) <Rith>.

After over a decade of intense work in many laboratories on the
$\gamma^{*} N \rightarrow \Delta$ reaction, the experiments have
reached a high  level of sensitivity  for the non-spherical
electric(E2) and Coulomb quadrupole(C2) amplitudes; these have
been observed with good precision as a function of $Q^{2}$ from
the photon point<Kotula>\cite{beck,blanpied} through 6 GeV$^{2}$ <
Sparveris, Smith, Ungaro> \cite{warren}- \cite{Ungaro}. For  the
highest $Q^{2}$  values it has been observed that the asymptotic
QCD predictions have not yet been reached<Ungaro>\cite{Ungaro}.
Lattice QCD <Alexandrou> \cite{Dina}with the appropriate chiral
extrapolations to the physical pion mass are in reasonable
agreement with experiment
<Pascalutsa-Vanderhaeghen>\cite{pasc,pvy}. However, due to the
high pion masses that are used in lattice calculations, and also
the limited accuracy of the present chiral extrapolations, the
theoretical errors are much larger than the experimental ones.
Therefore further theoretical work is required to make this
comparison quantitative.

For  the $\gamma^{*} N \rightarrow \Delta$ reaction at $Q^{2}
\leq 1 GeV^{2}$ the pion cloud is the dominant contributor to the
quadrupole amplitudes. This first became clear when  quark model
calculations for the quadrupole amplitudes were shown to be at
least an order of magnitude too small and even have the wrong
sign< Sparveris, Smith> \cite{sparveris,stave,sparveris2}. On the
other hand effective chiral field theory<Gail-Hemmert,
Pascalutsa-Vanderhaeghen>\cite{GHKP,gh,pasc,pvy}, and dynamic
model calculations<Drechsel-Tiator> \cite{sato_lee,dmt} which
include the effects of the pion-cloud are in approximate
agreement with experiment. This is expected due to the
spontaneous hiding (breaking) of chiral symmetry in QCD and  the
resulting, long range (low $Q^{2}$), effects of the pion-cloud.

In recent years there has been considerable progress in the
measurement of nucleon form factors by electron scattering
including polarization degrees of freedom <de~Jager>. These
include polarized electron scattering from polarized targets and
also detection of recoil nucleon polarization. There has also
been excellent progress in the measurement of virtual Compton
scattering. The results for nucleon form factors<de~Jager> and
virtual Compton scattering <Miskimen>\cite{VCS} experiments
indicate that  the pion-cloud is playing a significant role in
nucleon structure. It has been demonstrated that the concept of the
pion cloud is very interesting qualitatively but it cannot be
quantified in a model independent way <Meissner> \cite{disp-pion}.

There has been a considerable amount of work to improve the
extraction of the quadrupole amplitudes from the experimental
cross sections in the $ep \rightarrow \pi N$ reaction in the
$\Delta$ region. Empirical extractions of the multipole
amplitudes have considerably improved due to the increasing
sophistication and accuracy of both the experiments and the
reaction models <Drechsel-Tiator, Arndt>\cite{sato_lee,dmt, maid,
said}. A systematic comparison of results obtained by the
different models indicates that the model errors are {comparable
to the experimental errors
<Stave>\cite{sparveris,stave,sparveris2}. However, discriminating
between successful competing models or theoretical approaches,
which is necessary in order to  allow an understanding of the
detailed mechanisms that generate non spherical components
(amplitudes) in the hadron wavefunctions, requires even more
accurate experimental data and more precise analysis tools.
Towards this direction, the new method (AMIAS) for extracting
model independent multipoles from experiment which was presented
for the first time at this workshop <Stiliaris-Papanicolas>, is an
important first development.

Last, but not least, we believe that we can look forward to
exciting new developments in the determination of non-zero
angular momentum studies of hadrons, particularly the proton.
These will most likely include a new generation of high energy
probing as well as further refinements in the $\gamma^{*} N
\rightarrow \Delta$ reaction.



\begin{theacknowledgments}
We  acknowledge the help we received from  many colleagues who to
varying degree helped in the preparation of this manuscript; this
list certainly includes a good fraction of the hundred or so
physicists that participated in the two (Athens and MIT)
workshops. We are particularly indebted to C. Alexandrou, V.
Brown, U-G. Meissner, N. Sparveris S. Stave, E. Stiliaris and M. Vanderhaegen
for the help and the critical comments we received from them.
Finally we would like to thank Gerry Miller for suggesting the
title "Shape of Hadrons" for the second workshop, which we
adopted as the title of this volume.

    We would also like to thank the hosts of the two workshops.
For the 2006 workshop  the Institute of Accelerating Systems and Applications (IASA) and the National and Capodistrian University of Athens (NCUA). For the 2004 workshop the Laboratory for Nuclear Science and Bates Linear  Accelerator Center of the Massachusetts Institute of Technology (MIT).
\end{theacknowledgments}



\bibliographystyle{aipproc}   


\IfFileExists{\jobname.bbl}{} {\typeout{}
  \typeout{******************************************}
  \typeout{** Please run "bibtex \jobname" to optain}
  \typeout{** the bibliography and then re-run LaTeX}
  \typeout{** twice to fix the references!}
  \typeout{******************************************}
  \typeout{}
}

\end{document}